\documentstyle[amssymb,prc,aps]{revtex}

\begin{document}
\title{Optical model potentials involving loosely bound p-shell nuclei around 10
MeV/A}
\author{L. Trache}
\author{A. Azhari}
\author{H. L. Clark}
\author{C. A. Gagliardi}
\author{Y.-W. Lui}
\author{A. M. Mukhamedzhanov}
\author{R. E. Tribble}
\address{Cyclotron Institute, Texas A\&M University, College Station, \\
TX-77843, USA}
\author{F. Carstoiu}
\address{Institute of Atomic Physics, P. O. Box MG-6, Bucharest, Romania}
\date{\today}
\maketitle

\begin{abstract}
We present the results of a search for optical model potentials for use in
the description of elastic scattering and transfer reactions involving
stable and radioactive p-shell nuclei. This was done in connection with our
program to use transfer reactions to obtain data for nuclear astrophysics,
in particular for the determination of the astrophysical S$_{17}$ factor for 
$^{7}$Be(p$,\gamma )^{8}$B using two $(^{7}$Be$,^{8}$B$)$ proton transfer
reactions. Elastic scattering was measured using $^{7}$Li, $^{10}$B, $^{13}$%
C and $^{14}$N projectiles on $^{9}$Be and $^{13}$C targets at or about
E/A=10 MeV/nucleon. Woods-Saxon type optical model potentials were extracted
and are compared with potentials obtained from a microscopic double folding
model. Several nucleon-nucleon effective interactions were used: M3Y with
zero range and finite range exchange term, two density dependent versions of
M3Y and the effective interaction of Jeukenne, Lejeune and Mahaux. We find
that the latter one, which has an independent imaginary part, gives the best
description. Furthermore, we find the renormalization constant for the real
part of the folding potential to be nearly independent of the
projectile-target combination at this energy and that no renormalization is
needed for the imaginary part. From this analysis, we are able to eliminate
an ambiguity in optical model parameters and thus better determine the
Asymptotic Normalization Coefficient for $^{10}$B$\rightarrow ^{9}$B+p.
Finally we use these results to find optical model potentials for unstable
nuclei with emphasis on the reliability of the description they provide for
peripheral proton transfer reactions. We discuss the uncertainty introduced
by the procedure in the prediction of the DWBA cross sections for the ($^{7}$%
Be,$^{8}$B) reactions used in extracting the astrophysical factor S$_{17}(0)$%
.
\end{abstract}


\section{Introduction}

\label{intro} Transfer reactions have been proposed as an indirect method to
determine direct capture reaction rates at stellar temperatures for some
time \cite{Xu94,Ga95,Ros95}. Recently we used the Asymptotic Normalization
Coefficient (ANC) method to determine the cross section for the radiative
proton capture process $^{7}${Be}(p,$\gamma )$$^{8}${B} at solar energies,
or equivalently, the astrophysical factor, S$_{17}(0)$. The method relies on
the fact that at low energies a capture reaction to a loosely bound state is
a surface process. Its cross section is determined by the tail of the radial
overlap integral between the bound state wave function of the final nucleus
and those of the initial colliding nuclei. This overlap integral is
asymptotically proportional to a well known Whittaker function, and
therefore the knowledge of its asymptotic normalization alone determines the
cross section. This asymptotic normalization, in turn, can be determined
from the measurement of a transfer reaction involving the same vertex,
provided that this second reaction is also peripheral. In particular, we
determined S$_{17}(0)$ from measurements of the ANC for the $^{8}${B}$%
\rightarrow $$^{7}${Be} + p system utilizing the proton transfer reactions $%
^{10}${B}($^{7}${Be},$^{8}${B})$^{9}${Be} \cite{aa99} and $^{14}${N}($^{7}${%
Be}, $^{8}${B})$^{13}${C} \cite{aa99b}, at energies where the proton
transfer process is peripheral. Determining the ANCs from transfer reactions
involves distorted wave Born approximation (DWBA) calculations, and
therefore good, reliable optical potentials are needed. In particular, good
optical model potentials are needed in both the initial and the final
channels involving $^{7}$Be and $^{8}$B radioactive nuclei in each reaction
in order to compute the DWBA proton transfer cross sections. Because elastic
scattering data with these projectiles cannot easily be obtained and/or are
not precise enough to extract reliable and unambiguous optical potentials,
we have studied elastic scattering for several combinations of p-shell
nuclei at energies close to those appearing in the reactions of interest. We
then determine procedures to extract optical model potentials for the cases
involving radioactive partners.

Angular distributions up to the nuclear rainbow angle were measured in seven
experiments using $^{7}$Li, $^{10}$B, $^{13}$C and $^{14}$N projectiles on $%
^{9}$Be and $^{13}$C targets at bombarding energies at or around E/A=10
MeV/nucleon. They were fitted with phenomenological potentials with volume
Woods-Saxon real and imaginary terms. The phenomenological optical model
potentials found for all systems were then compared with the potentials
calculated with microscopic double folding procedures, using six effective
nucleon-nucleon interactions: M3Y with zero range and finite exchange term,
the density dependent M3Y interaction in the forms extracted recently by
Khoa et al. \cite{khoa} (BDM3Y1, BDM3Y3) and the interaction of Jeukenne,
Lejeune and Mahaux (JLM) \cite{jlm} in two versions. For the calculations,
one-body densities were obtained in a standard spherical Hartree-Fock
calculation using the density functional of Beiner and Lombard \cite{bei},
with a slight modification of the surface term in order to fit the
experimental binding energy for each nucleus. These densities were used in
the double folding procedure. The renormalization coefficients needed for
the analysis of elastic data with these double folding potentials are
extracted and discussed here.

In two cases the elastic scattering data were also used to extract the
parameters of the optical model potentials that were needed for DWBA
calculations to determine the ANCs for the $^{10}${B}$\rightarrow $$^{9}${Be}
+ p and $^{14}$N $\rightarrow ^{13}$C+ p systems from the $^{9}$Be($^{10}$B,$%
^{9}$Be)$^{10}$B \cite{ak97} and $^{13}$C($^{14}$N,$^{13}$C)$^{14}$N \cite
{tra98} reactions, respectively, and were included in those publications.
These measurements also allowed us to test the results of our folding
procedures for the proton transfer reactions by comparing the results of the
DWBA calculations that we obtain from the double folding model potentials
with those we obtain using the phenomenological potentials. From the
analysis presented below, we are able to eliminate the uncertainty in the
optical potentials found in Ref. \cite{ak97}. This results in a better
determination of the ANC for $^{10}$B$\rightarrow ^{9}$Be+p. We chose the
elastic scattering of $^{7}$Li on both targets as being close to what we
expect for the scattering of $^{7}$Be. Both $^{7}$Li and $^{7}$Be nuclei
have low binding energies and large break-up probabilities. The elastic
scattering of $^{13}$C on the $^{9}$Be target was studied as this is close
to the exit channel $^{13}$C+$^{8}$B of the second proton transfer reaction $%
^{14}$N($^{7}$Be,$^{8}$B)$^{13}$C.

The double folding procedure and the parameters extracted for the effective
nucleon-nucleon interaction from the present experimental data were also
checked for other projectile-target combinations in the same mass and energy
region for which data exist in the literature. The results were in agreement
with those found from the study of the seven cases described here. Using the
techniques developed here, we calculated the optical model parameters
required to extract the ANC for the $^{7}$Be+p$\rightarrow ^{8}$B system,
and consequently S$_{17}$(0), from the studies of the $^{10}$B($^{7}$Be,$%
^{8} $B)$^{9}$Be \cite{aa99} and $^{14}$N($^{7}$Be,$^{8}$B)$^{13}$C \cite
{aa99b} reactions. Section \ref{sec2} presents the experiments and the
procedures used in the data reduction. We extract the Woods-Saxon optical
model potentials from fits to the elastic scattering data in Section \ref
{sec3}, and compare them with those microscopically calculated in Section 
\ref{sec4}. In sections \ref{sec5} and \ref{sec6}, we describe a global
optical potential for interactions involving loosely bound p-shell nuclei
and its implications for the $^{10}$B($^{7}$Be,$^{8}$B)$^{9}$Be and $^{14}$N(%
$^{7}$Be,$^{8}$B)$^{13}$C reactions. Finally, Section \ref{concl} presents
the conclusions.

\section{The experiments}

\label{sec2}

The experiments were carried out using beams from the Texas A\&M University
K500 superconducting cyclotron. A list of the measurements is given in Table 
\ref{tab1}. The experimental setup and the data reduction procedures were
similar to those already described in Ref. \cite{ak97}. The
Multipole-Dipole-Multipole (MDM) magnetic spectrometer \cite{Oxs} was used
to analyze the scattered particles and the reaction products. The beams were
prepared using the beam analysis system \cite{BAS}, which allows for the
control of the energy and angular spread of the beam. Self-supported $^{9}${%
Be} and $^{13}$C targets, about 200-300 $\mu $g/cm$^{2}$ thick, obtained by
evaporation, were placed perpendicular to the beam in the sliding-seal
target chamber of the MDM. The magnetic field of the spectrometer was set to
transport fully stripped ions to its focal plane, where they were observed
in the modified Oxford detector \cite{Oxd}. There, the position of the
particles along the dispersive direction was measured with resistive wires
at four different depths within the detector, separated by about 16 cm each.
For particle identification we used the specific energy loss measured in the
ionization chamber and the residual energy measured in a NE102A plastic
scintillator located in air, just behind the output window of the detector.
The input and output windows of the detector were made of 1.8 and 7.2 mg/cm$%
^{2}$ thick Kapton foils, respectively. The ionization chamber was filled
with pure isobutane at pressures between $30-50$ torr. The entire horizontal
acceptance of the spectrometer, $\Delta \theta =\pm 2^{\circ }$, and a
restricted vertical opening, $\Delta \phi =\pm 0.5^{\circ }$, were used in
these measurements. Raytracing was used to reconstruct the scattering angle
in the analysis of the data. For this purpose, in addition to RAYTRACE \cite
{RAY} calculations, angle calibration data were obtained at several angles
by using an angle mask consisting of five openings of $\delta \theta $=0.1$%
^{\circ },$ located at -1.6$^{\circ }$, -0.8$^{\circ }$, 0$^{\circ }$, +0.8$%
^{\circ }$ and +1.6$^{\circ }$ relative to the central angle of the
spectrometer. Moving the spectrometer from $\theta _{lab}=4^{\circ }$ to $%
54^{\circ }$ we covered the angular ranges listed in Table \ref{tab1}.
Typically we moved the spectrometer by $2^{\circ }$ or $3^{\circ }$at a
time, allowing for an angle overlap that provided a self-consistency check
of the data at all angles. Normalization of the data was done using current
integration in a Faraday cup. Focal plane reconstruction was done at each
angle using the position measured with the signals in the wire nearest to
the focal plane and using the detector angle obtained from the position
measured at two of the four wires. The angular range of $\Delta \theta
=4^{\circ }$ covered by the acceptance slit was divided into 8 bins,
resulting in 8 points in the angular distribution being measured
simultaneously, with each integrating over $\delta \theta _{lab}=0.5^{\circ
} $.

The measurements with the angle mask showed that the resolution in the
scattering angle (laboratory) was $\Delta \theta _{res}=0.20^{\circ
}-0.25^{\circ }$ full-width at half maximum (FWHM). This includes a
contribution from the angular spread of the beam of about $0.1^{\circ }$.
The best energy resolution obtained at forward angles was 150 keV FWHM for $%
^{10}$B on $^{9}$Be, 230 keV for $^{14}$N on $^{13}$C and 150 and 220 keV
for the scattering of $^{7}$Li at 63 MeV and 130 MeV, respectively. It
degraded as we advanced to larger angles due to the large kinematic factor, $%
k=\frac{1}{p}\frac{dp}{d\theta }$, coupled with the finite angular spread in
the beam. However, it never degraded our ability to isolate the elastic
peak, even in the most demanding case of the $^{7}$Li experiments where the
first excited state of the projectile is only 477 keV away. A sample
spectrum taken in one of these most demanding cases, $^{7}$Li (130 MeV) + $%
^{9}$Be, is presented in Figure 1 where the good separation is clear. The
active length of the focal plane allowed us to cover a total excitation
energy of about 7 MeV, centered around the elastic peak. Thus we were able
to measure inelastic scattering to the lowest excited states of the
projectile-target systems at the same time. These inelastic scattering data
were used as additional information to check the experimental procedures,
and in a few cases we compared the inelastic transition strength obtained in
these experiments (deformation parameters or deformation lengths) with those
available in literature.

During the experiments, particular emphasis was placed on obtaining accurate
absolute values for the cross sections, and therefore target thickness and
charge collection factors were determined by a two-target method and by
normalization to Rutherford scattering at forward-most angles, as described
in \cite{ak97}. Combining the results of these independent determinations,
we conclude that we have an overall normalization accuracy of 7\% for the
absolute values of the cross sections for both the elastic scattering $^{9}${%
Be}($^{10}${B},$^{10}${B})$^{9}${Be} data and for the proton transfer $^{9}${%
Be}($^{10}${B},$^{9}${Be})$^{10}${B} data \cite{ak97} and for the elastic
scattering $^{13}$C($^{14}$N,$^{14}$N)$^{13}$C and proton transfer reaction $%
^{13}$C($^{14}$N,$^{13}$C)$^{14}$N \cite{tra98}. The normalization for the
absolute values of the cross section was made in the $^{13}$C\ (130 MeV) + $%
^{9}$Be case only using the nine most forward angle data points in the
angular distribution and is estimated to be accurate to 10\%. For the
experiments with the $^{7}$Li beam, we also determined the target thickness
by measuring the energy loss of alpha particles from a $^{228}$Th source and
the accuracy in normalization is 9\%.

\section{Phenomenological Analysis}

\label{sec3}

The elastic scattering data have been fitted using the code OPTIMINIX \cite
{flo} in a standard optical model analysis using volume Woods-Saxon form
factors with the standard notation: 
\begin{equation}
{\rm U(r)}=-\left( {\rm Vf_{V}(r)+iWf_{W}(r)}\right)  \label{p1}
\end{equation}
where 
\begin{equation}
{\rm f_{x}(r)=\left[ 1+\exp \left( \frac{r-r_{x}(A_{1}^{1/3}+A_{2}^{1/3})}{%
a_{x}}\right) \right] ^{-1}}  \label{p2}
\end{equation}
and x=V,W stands for the real and imaginary parts of the potentials,
respectively. Fits using the codes PTOLEMY \cite{PTO} and ECIS \cite{ECIS}
gave similar results. Only the central components have been included in the
optical potential, since vector and higher rank tensor spin-orbit couplings
have little or no influence on the cross sections.

In Figure \ref{fig2} we present the angular distributions measured for $%
^{10} $B+$^{9}$Be at E($^{10}$B)=100 MeV, $^{14}$N+$^{13}$C at E($^{14}$%
N)=162 MeV and $^{13}$C+$^{9}$Be at E($^{13}$C)=130 MeV and in Figure \ref
{fig3} those for $^{7}$Li+$^{9}$Be,$^{13}$C at E($^{7}$Li)=63 and 130 MeV.
All angular distributions display typical patterns for elastic scattering,
dominated by strong absorption with Fraunhoffer oscillations of large
amplitude around the crossing point, followed by less developed structures
at larger angles. The Fraunhoffer oscillations are expected for such
systems, due to the small Sommerfeld parameter $\eta \simeq 1$. The curves
are the fits to the data. Inspecting Figures \ref{fig2} and \ref{fig3}, one
observes that all potentials predict somewhat deeper minima than the data
show. This effect is partially attributable to the finite angular
resolution. The rest is probably due to the incoherent contribution of the
substantial quadrupole moment of some of the partners (like $^{9}$Be).
During the normalization procedure, the theoretical curves were convoluted
with the experimental angular resolution and binning, using the code ECIS,
but this was found to have no influence on the fits.

The optical model parameters extracted for all seven cases studied here are
presented in Table \ref{tab2}. In addition to the depth, reduced radius and
diffuseness for the real and imaginary parts of the potential, the table
gives the chi-square for the fit ($\chi ^{2}$), the total reaction cross
section ($\sigma _{R}$), the values of the volume integrals per pair of
interacting nucleons for the real (J$_{V}$) and imaginary parts (J$_{W}$) of
the potential, respectively, and the root-mean-square radii of the real (R$%
_{V}$) and imaginary (R$_{W}$) potentials. The parameters were obtained by
griding the initial strength of the real potential in small steps in the
range V=50-250 MeV in order to identify the local minima and then searching
for minima on all parameters with no constraints. In this way 3-4 families
of potentials have been found for each case. Usually, a characteristic jump
of 50-70 MeV fm$^{3}$ in the volume integral of the real part of the optical
potential serves to identify these potentials as discrete members of a
sequence of potentials which give a comparable description of the data. The
members of each family of potentials are connected by the well known
continuous Igo ambiguity: $Vexp(R_{V}/a_{V})=const.$ \cite{Igo}. This arises
since, due to the strong absorption, the cross section is sensitive only to
the tail of the potential. The Igo potentials of the same family have
practically the same volume integral and the same radius and therefore the
discrete families of potentials can be identified by the values of these
parameters. The absorption is seen to be independent of the strength and
shape of the real part of the optical potential and, as a consequence, the
reaction cross section is more or less constant along the sequence in each
case. Also we notice that generally the radii of the imaginary potentials
are about 20\% larger than those for the real potentials, in agreement with
previous observations \cite{sat}.

For the $^{10}$B+$^{9}$Be case, it appears that potential 1, which has the
smallest chi-square, provides the most realistic description, and potential
3 can be rejected. In the angular range covered, the prediction of potential
2 for the elastic scattering differs from that of potential 1 primarily in
the depths of its minima. We reached the same conclusion from the comparison
of the DWBA calculations for the proton transfer reaction $^{9}$Be($^{10}$B,$%
^{9}$Be)$^{10}$B studied in the same experiment: whereas potentials 1 and 2
give a very good description of the shape of the angular distribution and
similar absolute values, potential 3 predicts a reaction cross section which
is far too small \cite{ak97}. To further clarify the features of the angular
distribution we have performed a near-far decomposition of the scattering
amplitude, with one potential taken as the reference potential. Shown in
Figure \ref{fig4}a are the cross-sections due to the near-side and far-side
components of the total scattering amplitude. Around the crossing angle of $%
\theta _{c.m.}=16^{\circ }$, the strong interference between the near- and
far-side components results in Fraunhoffer oscillations of large amplitude.
Beyond this angle, the near-side component drops rapidly and the angular
distribution is dominated by the far-side component, which falls off
smoothly. No significant structure could be identified in this region. This
structureless behavior can be understood qualitatively in terms of the
transparency for the low partial waves implied by the refractive potential.
In the case of $^{10}$B + $^{9}$Be, the rainbow partial wave has $l_{R}=16$
and the associated scattering matrix elements are of the order $%
|S_{l}|\simeq 3\cdot 10^{-3}$ for $l<<l_{R}$. Thus, the refractive nature of
the potential is sufficient to allow the interference between waves with $%
l<16$ and higher ones to produce the smooth behavior. Comparison of the
potential elastic scattering branch (forward angles) and of the elastic
proton exchange branch (backward angles) in Figure 3 of Ref. \cite{ak97}
shows that the interference between these two mechanisms has no sizable
effect in the angle range considered here and was not considered in the
analysis. The data for the $^{13}$C (130 MeV) + $^{9}$Be experiment show
similar features, with 3 families of potentials found.

For the $^{14}${N} + $^{13}${C system the }volume integral and radius for
the absorptive part of the optical potential seems to be independent of the
real potential, resulting in a constant total reaction cross section along
the sequence with an average value of $1535$ mb. The optical model total
reaction cross section is consistent with the experimental value measured by
DiGregorio {\it et al.} at $161.3~$MeV, $\sigma =1463\pm 100$ mb \cite{grego}%
. All potentials give reasonable $\chi ^{2}$, but potential 1 listed in the
table gives the smallest value and is the only one that fits the data at the
largest angles. This potential has a volume integral per pair of interacting
nucleons close to that which we found for $^{10}$B$+^{9}$Be elastic
scattering at similar velocities. Potential 1 was adopted for the DWBA
calculation of the proton exchange process $^{13}$C($^{14}$N,$^{13}$C)$^{14}$%
N as described in Ref. \cite{tra98}, while the others were used to estimate
the uncertainty due to the choice of optical model parameters. Similar
insight on the relative role played by the refractive and absorptive parts
of the optical potential may be obtained from the far side-near side
decomposition of the scattering amplitude corresponding to potential 1 which
is presented in Figure \ref{fig4}b. The far-side component is represented by
the dashed line and the near-side component by the dotted line and their
coherent sum by the continuous line. For angles around the crossing where
the two components have comparable amplitude and strongly interfere, a
typical Fraunhofer diffraction pattern emerges with large amplitude
oscillations equally spaced by $\Delta \theta =\pi /l_{g}=\pi /30$, where $%
l_{g}$ is the grazing angular momentum. Beyond this angle, the near side
component is completely damped by the strong absorption and we are left with
the far side exponential tail that is characteristic of far-side dominance.
No significant structure could be identified up to the nuclear rainbow
angle, which in this case is $\theta _{R}=83^{\circ }$. Similar to the case
of the $^{10}$B+$^{9}$Be experiment, our measurements show that we do not
have interference effects between the potential scattering predominant at
forward angles and the elastic proton exchange predominant at backward
angles (see Figure 2 of Ref. \cite{tra98}) in the angular range considered.

The potentials found in the phenomenological WS analysis of $^{7}$Li
scattering are given in Table \ref{tab2}. A similar result emerges, with
discrete ambiguities represented by up to 4 families found in each case.
Similar values are found for the volume integrals for the real and imaginary
parts as for the rest of the cases studied above. We notice however that the
reduced radii r$_{V,W}$ are small and the diffusivities $a_{V,W}$ of the
potentials are unusually large. This agrees with findings in other analyses
for such light systems. Figure \ref{fig4}c shows the far-side, near-side
decomposition for the $^{7}$Li+$^{13}$C system at 63 MeV, with conclusions
similar with those for the cases discussed above.

It is interesting to note that, for all but one of the cases shown in Table 
\ref{tab2}, the first of the potentials always has a similar volume integral
for the real part: J$_{V}\approx $ 220 MeV fm$^{3}$, and that the imaginary
potentials are independent of the real part, predicting the same total cross
sections.

As mentioned above, the spin dependent components of the optical potential
have been omitted. In the absence of any polarization data, exploratory
calculations for the $^{10}$B + $^{9}$Be system, using the same vector
spin-orbit term as for $^{6}$Li + $^{12}$C \cite{coo} at E=156 MeV, did not
result in any noticeable effects on the elastic cross section in the
measured angular range. For several of the cases studied here, we also did a
Fourier-Bessel analysis of the data, similar to that in Ref. \cite{GRF84},
and found that the phenomenological Woods-Saxon shapes assumed in Eq. 2 are
adequate.

\section{Folding Model Analysis}

\label{sec4}

In addition to the analysis with Woods-Saxon type potentials, the data have
been analyzed in the framework of a semi-microscopic folding model. We
followed a Hartree-Fock procedure to obtain the densities in the two
partners, then used double folding with known nucleon-nucleon interactions.
The wave functions and the densities for all nuclei involved were obtained
in a standard spherical Hartree-Fock calculation using the energy density
functional of Beiner and Lombard \cite{bei}. This functional describes
nuclear matter and the bulk properties of finite nuclei well. In the
calculations, the parameters of the surface terms were adjusted slightly in
order to reproduce the experimental total binding energy. This is an
important constraint on the calculation, especially for nuclei with small
separation energies such as $^{9}$Be and $^{7}$Li. Usually this correction
amounts to a few percent with respect to the original parameters and
substantially improves the description of the single particle levels close
to the Fermi level. The calculated binding energies and the rms radii that
were obtained are given in Table \ref{tab_dens} and compared with the
experimental ones. A similar procedure has been used by Hoshino et al. \cite
{hos} to describe the structure of the $^{11}$Be nucleus.

In the double folding procedure, we used a number of G-matrix effective
nucleon-nucleon (NN) interactions. The first one is the well known M3Y
interaction. Recall that the nucleus-nucleus potential in the double folding
model is given by 
\begin{equation}
V_{fold}(R)=\int d\vec{r}_{1}d\vec{r}_{2}\rho _{1}(r_{1})\rho
_{2}(r_{2})v_{eff}(\vec{r}_{1}+\vec{R}-\vec{r}_{2})
\end{equation}
where $\rho _{1,2}$ are the single particle densities, and the interaction
operator is of the form 
\[
v_{eff}(r)=v_{D}(r)+P_{1,2}^{ex}v_{ex}(r) 
\]
where the direct and exchange parts are averaged over spin-isospin states
and $P_{1,2}^{ex}$ is the knock-on exchange operator in coordinate space. We
assumed, as usual, that the one nucleon exchange knock-on term, which
involves the exchange between the interacting nucleons, is dominant with
respect to all other exchange contributions. The parameters for the direct
and exchange components of M3Y were taken from Ref.\cite{bert}. In the
standard version \cite{sat}, the isoscalar component of the interaction
consists of a finite range direct term, supplemented by an energy dependent
zero range pseudo-potential which simulates well the one nucleon knock-on
contribution to the interaction. The small isovector component of the
interaction has also been included in the calculation and the corresponding
results are denoted by M3Y/ZR throughout the paper. A finite range version
of the M3Y interaction was also used for some of the systems analyzed in
this paper. The lack of any explicit density dependence in the effective
interaction results in potentials that are too deep in the interior to
reproduce correctly the rainbow features at large angles observed e.g. in
alpha scattering at higher energy\cite{satch}. This can be corrected by
making the effective NN interaction depend upon the density of the nuclear
matter in which the interacting nucleons are immersed. The requirement that
nuclear matter saturate ensures that this density dependence reduces the
strength of the interaction as the density increases, weakening the folding
potential in the interior while leaving the surface values practically
unchanged. For our purpose we adopted more recent interactions called
BDM3Yn(n=1 and 3) which have been shown to give a good description of light
ion scattering in a wide range of incident energies \cite{khoa}. These
interactions are based upon a G matrix derived from the Reid soft-core NN
potential. They incorporate a linear (n=1) or cubic (n=3) density dependence
with parameters adjusted to give saturation in nuclear matter at the correct
density and binding energy. The two interactions give very different
compressibilities for nuclear matter ($K=230$ MeV for$~n=1$ and $K=475$ MeV
for$~n=3)$ covering a broad range of equations of state. We note that at
present $K_{\infty }=231\pm 5$ MeV has experimental support \cite{you99},
which would indicate a preference for BDM3Y1 in the description of heavy ion
elastic scattering.

Also, we have used the G-matrix interaction of Jeukenne, Lejeune and Mahaux
(JLM)\cite{jlm}, which is obtained in a Brueckner-Hartree-Fock (BHF)
approximation from the Reid soft-core nucleon-nucleon potential. This
interaction is complex, energy and density-dependent and, therefore,
provides simultaneously both real and imaginary parts of the optical
potential. The interaction has been considered recently by Bauge, Delaroche
and Girod \cite{bauge} in an extensive study of nucleon scattering on a wide
range of target masses and incident energies. Some shortcomings of the
original interaction were also corrected in Ref. \cite{bauge}. For
completeness, we describe below the main steps in the derivation of our
potentials, taking into account the improvements recommended in Ref. \cite
{bauge}.

The optical potential for a nucleon of energy $E$ traversing nuclear matter
of density $\rho $ is written as: 
\begin{equation}
U_{NM}(\rho ,E)=V_{0}(\rho ,E)+\alpha \tau V_{1}(\rho ,E)+i[W_{0}(\rho
,E)+\alpha \tau W_{1}(\rho ,E)]
\end{equation}
where $\alpha =(\rho _{n}-\rho _{p})/(\rho _{n}+\rho _{p})$ and $\tau =\pm 1$
for neutrons and protons, respectively. Explicit expressions for various
terms are : 
\begin{equation}
V_{0}(\rho ,E)=\sum_{i,j=1}^{3}a_{ij}\rho ^{i}E^{j-1}
\end{equation}
\begin{equation}
W_{0}(\rho ,E)=\left( 1+{\frac{{D}}{{(E-\epsilon _{F}(\rho ))^{2}}}}\right)
^{-1}\sum_{i,j=1}^{4}d_{ij}\rho ^{i}E^{j-1}
\end{equation}
\begin{equation}
V_{1}(\rho ,E)={\frac{{\tilde{m}}}{{m}}}\Re N(\rho ,E)
\end{equation}
\begin{equation}
W_{1}(\rho ,E)={\frac{{m}}{{\bar{m}}}}\Im N(\rho ,E)
\end{equation}
The matrix coefficients $a_{ij}$, $d_{ij}$, the Fermi energy $\epsilon _{F}$
and the BHF expression of the auxiliary function $N(\rho ,E)$ are given in 
\cite{bauge}. The quantities $\tilde{m}/m$ and $\bar{m}/m$ are the $k$ mass
and the $E$ mass, respectively, and represent a measure of the true
nonlocality and the true energy dependence of the optical potential.

In applications for heavy ions we interpret the quantity 
\begin{equation}
v_{0}(\rho ,E)=(V_{0}(\rho ,E)+iW_{0}(\rho ,E))/\rho
\end{equation}
as the (complex) isoscalar, density- and energy-dependent NN effective
interaction. The heavy ion potential is given then by the folding integral: 
\begin{equation}
V(R)=\int d\vec{r_{1}}d\vec{r_{2}}\rho _{1}(r_{1})\rho _{2}(r_{2})v_{0}(\rho
,E)\delta (\vec{s})
\end{equation}
with $\vec{s}=\vec{r_{1}}+\vec{R}-\vec{r_{2}}$. Similarly, the quantity 
\begin{equation}
v_{1}(\rho ,E)=(V_{1}(\rho ,E)+iW_{1}(\rho ,E))/\rho
\end{equation}
is interpreted as the (complex) isovector, density- and energy-dependent NN
effective interaction. The corresponding heavy ion potentials are obtained
from a folding integral similar to that in Eq. (10), replacing $v_{0}$ by $%
v_{1}$ and the single particle densities $\rho _{1,2}$ by the isovector
densities $(\rho _{n}-\rho _{p})_{1,2}$. Usually such terms have little
influence in the total optical potential because the isovector densities are
small for normal nuclei in the p-shell; however we have included them in the
analysis since such terms can have some importance in the case of loosely
bound nuclei with very different proton and neutron single particle
densities. Two approximations for the local density have been used. The
first of them reads: 
\begin{equation}
\rho =\left( \rho _{1}(\vec{r_{1}}+{\frac{{\vec{s}}}{{2}}})\rho _{2}(\vec{r}%
_{2}-{\frac{{\vec{s}}}{{2}}})\right) ^{1/2}
\end{equation}
which amounts to an estimate of the local density as the geometric average
of the individual single particle densities, each of them evaluated at the
mid distance between the interacting nucleons. This approximation has been
used by Campi and Sprung \cite{campi} in Hartree-Fock calculations with
density-dependent forces. With this approximation, the local density never
exceeds the saturation value for nuclear matter density $\rho _{0}$. We
remind the reader that the JLM effective interaction is defined only for
density values satisfying $\rho \le \rho _{0}$. Potentials obtained with the
above approximation are labeled below as JLM(1). The second approximation
for the local density uses the arithmetic average of the individual
densities: 
\begin{equation}
\rho ={\frac{{1}}{{2}}}\left( \rho _{1}(\vec{r_{1}}+{\frac{{\vec{s}}}{{2}}}%
)+\rho _{2}(\vec{r}_{2}-{\frac{{\vec{s}}}{{2}}})\right)
\end{equation}
A similar approximation was used by the authors of Ref.\cite{kob} in their
derivation of the density-dependent version of M3Y, except for the factor of
one-half in front of the parentheses which is introduced here in order to be
consistent with the assumptions of the JLM model. Potentials calculated with
this approximation are denoted by JLM(2). It has been shown by the authors
of Refs. \cite{jlm} and \cite{bauge} that the local density approximation is
substantially improved by replacing the $\delta $ function in integrals of
the type (10) by finite range form factors of the gaussian shape: 
\begin{equation}
g(\vec{s})=\left( {\frac{{1}}{{t\sqrt{\pi }}}}\right) ^{3}e^{-s^{2}/t^{2}}
\end{equation}
Since the finite range form factors are normalized to one, the volume
integrals of the folding potentials are not affected, only the $rms$ radii
are increased, depending on the values one chooses for the range parameter $%
t $. Our phenomenological analysis shows clearly that the bulk of the
elastic scattering experimental data require larger radii for the absorptive
part of the heavy ion optical potentials as compared to the real part.
Extensive numerical calculations with both versions of the JLM interaction
showed that optimum values for the range parameters are $t_{R}=1.2~$fm and $%
t_{I}=1.75~$fm. A similar need for different radii of the imaginary and real
parts of the optical potential has been emphasized recently by Satchler and
Khoa \cite{satch}. Of course slightly improved fits could be obtained in
each individual case by varying also the range parameters around these
values. For example, our $^{7}$Li data were better fitted with a larger $%
t_{I}$. However, finding such variations for individual data sets goes
beyond the purpose of the present paper.

It is known that p-shell nuclei elastic scattering, some of which involve
loosely bound nuclei, cannot be described successfully without a substantial
renormalization of the folding form factor \cite{sat1}. The strong coupling
with breakup and neutron transfer channels is responsible for such an
effect. The usual procedure to simulate the repulsive effect of the real
part of the dynamic polarization potential \cite{sak} arising from such
coupling is to introduce a multiplicative constant for the real folding form
factor. In the folding model with ${\it {real}}$ effective interactions the
absorption is accounted for phenomenologically by adding an imaginary
potential of the same shape as the real part: 
\begin{equation}
{\rm U(r)}={({\rm N_{V}+iN_{W})}}V_{fold}(r)  \label{f1}
\end{equation}
whereas for the cases when the effective interaction also has an {\it %
imaginary} component, the renormalization is: 
\begin{equation}
{\rm U(r)}={\rm N_{V}}V_{fold}(r){\rm +iN_{W}}W_{fold}(r).  \label{f2}
\end{equation}
The resulting potentials differ from the Woods-Saxon shape at small
distances, but can be easily fitted with such forms in their surface region.
We reanalyzed all our elastic scattering data using double folding
potentials obtained with the six effective interactions outlined above. The
renormalization constants ${\rm N_{V}}$ and ${\rm N_{W}}$ were further
adjusted to fit the elastic scattering data using Eq. \ref{f1} in the case
of M3Y and BDM3Y forces and Eq. \ref{f2} for the two versions of the JLM
interactions. The results of the fits are shown in Figures \ref{fig5} and 
\ref{fig6}, and the parameters are displayed in Table \ref{tab4}. In
general, fits of reasonable quality were obtained with all interactions.
However, the JLM interaction not only gives the best fits as compared to the
other interactions, but also provides renormalization constants with minimal
dispersion for all projectile-target combinations considered. This indicates
that the mass dependence of the optical potential is properly taken into
account by these effective interactions through the density dependence. As a
rule, all folding potentials need a substantial renormalization for the real
part of the optical potential, emphasizing that the dynamic polarization
potential plays an important role for p-shell nuclei elastic scattering at
low energies. Density-dependent effects, such as those taken into account by
BDM3Y forces, lead only to a slight increase in the real normalization
constant ${\rm N_{V}}$ as compared to the original density-independent
interaction M3Y/ZR, suggesting a need for a stronger density dependence at
the potential surface. Inspecting Figures \ref{fig5} and \ref{fig6} one sees
that is hard to distinguish between the two versions of this force since
both of them give a comparable description of the data. This is likely a
consequence of the fact that the present data give information on the
optical potential in a limited spatial region centered around the strong
absorption radius where the two do not differ much.

\section{Extracting a global optical potential}

\label{sec5}The analysis done as described in the previous section leads us
to the conclusion that we can find a way to predict optical model potentials
with some reliability. As already noted before, the situation is complicated
by the fact that the nuclei involved are loosely bound and we expect to have
important effects from break-up channels. Satchler and Love \cite{sat}
concluded earlier that the renormalization of the real part of the double
folding potentials is considerable, particularly for loosely bound nuclei
where break-up is important. The energies studied here, around 10
MeV/nucleon, are known to lead to sizable effects due to the dynamic
polarization contribution to the optical potential \cite{satch}. This is
most likely the explanation behind the need for a substantially reduced real
well depth. The renormalization coefficients are presented in Table \ref
{tab4} and in Figure \ref{fig7} for both the real and imaginary part of the
potentials. If we compare the results for the same nucleon-nucleon
interaction, we see that similar renormalization constants are obtained for
all systems when at least one of the participating nuclei is weakly bound.
In particular when density-dependent effective interactions (JLM, BDM3Y1,
BDM3Y3) are used, the renormalization constants are very stable, with a
standard deviation of a few percent around the average value. This suggests
that one can indeed obtain the optical model potentials for pairs of
projectile-target nuclei for which data are not available, or are scarce, by
using a folding procedure to obtain the geometrical parameters and the
renormalization constants extracted above. Studies like the one comparing
the scattering of $^{6}${Li} and $^{11}${Li} lead to a similar conclusion,
and show that the energy dependence in the potential is smooth and rather
weak \cite{cars}. Furthermore, the renormalization factors that we find here
are comparable to those found for $^{6}${Li}+$^{12}${C} near this energy
when the M3Y and JLM interactions are used \cite{cars}. In a few cases the
finite range version of the exchange term in the M3Y interaction was also
checked but the results were not improved over those obtained with the zero
range version of it. Given our suspicion that the localization procedure
used to obtain these finite range calculations might not work properly in
very light nuclei, we do not discuss the results here, but they are included
in Table \ref{tab4}. From all six effective nucleon-nucleon interactions
used above, we favor the one denoted JLM(1) because it gives a slightly
better fit than the others and the renormalization coefficients have the
smallest spread around the average value (last rows in Table \ref{tab4}). In
contrast to the real potential, no renormalization is needed for the
imaginary part of the calculated potential, a sign that the imaginary part
of the effective interaction and its density dependence are well accounted
for. There might be a remaining slight dependence of the renormalization on
energy, as found in other studies, but our data are insufficient to extract
a definite conclusion on this dependence. However, it seems likely that most
of the energy dependence is taken care of by the energy dependence of the
effective interaction and by the density dependence used in the
calculations. We also checked our double folding procedure on other systems
than those mentioned above, and included the results in Table \ref{tab4}.
Whereas we obtain very good fits to the data over a large mass and energy
range, thus confirming the appropriateness of the JLM(1) effective
interaction and of the smear function and ranges used in Eq. 14, the
resulting renormalization coefficients, when using alpha particles for
example, differ from those for the p-shell nuclei studied here and point to
the conclusion that the present coefficients have only a local
applicability. Analysis of alpha scattering of up to 60 MeV/nucleon on
stable targets, lead to renormalization coefficients for the real part about
a factor two larger. This is surely a reflection of the differences between
the very well bound $^{4}$He nuclei and the loosely bound partners studied
here. In order to obtain more complete information on the renormalization
constants, we have included in our analysis two angular distributions
involving the elastic scattering of another loosely bound p-shell nucleus $%
^{6}$Li on light targets at 16 MeV/A \cite{schw} and $^{7}$Li+$^{12}$C at
two energies \cite{zel80}. The volume integrals of the renormalized double
folded potentials agree with the volume integrals of the first of the
phenomenological potentials found and suggest that the phenomenological
potentials with J$_{V}\thickapprox $ 220 MeV fm$^{3}$ give the most
realistic description.

Data in Figure \ref{fig7} and Table \ref{tab4} show that the renormalization
coefficient of the real potential calculated with the JLM(1) interaction is
somewhat higher for $^{14}$N+$^{13}$C than the average of the remaining 6
cases measured here. This is the only projectile-target combination where
both nuclei are well bound, thus we should expect a smaller contribution
from the polarization potential. The averages and standard deviations for
all 7 cases are ${\rm N_{V}}$=0.378$\pm $0.034 (or 9\%) and ${\rm N_{W}}$%
=1.004$\pm $0.135 (13\%), respectively. Excluding the $^{14}$N+$^{13}$C
system we find the averages ${\rm N_{V}}$=0.366$\pm $0.014 (or 4\%) and $%
{\rm N_{W}}$=1.000$\pm $0.087 (9\%). We see that the value of the
renormalization coefficient is very stable. We suspect that a large part of
the spread around the average of the renormalization coefficient for the
imaginary potential comes from the uncertainties in the absolute
normalization of our data. The real part of the potential, which is fixed
mostly by the position of the oscillations in the angular distributions, is
less sensitive to this absolute normalization. This is the reason we exclude
the data of other groups (lower part of Table \ref{tab4}) from the present
averaging procedure (last row in Table \ref{tab4}). We note that,
deformation, which is important in some p-shell nuclei, is not included in
any way in our calculations, due to the use of spherical Hartree-Fock
density distributions.

Further, we checked to see to what extent the double folding potentials,
renormalized to fit the elastic scattering data, give the same results as
the phenomenological potentials when used to calculate the cross sections
for the proton transfer reactions. For the reaction $^{9}$Be($^{10}$B,$^{9}$%
Be)$^{10}$B we found that the cross section calculated with the JLM(1)
potential, renormalized as above, differs by less than 1\% (integral over
the angles from 0$^{\circ }$ to 45$^{\circ }$ ) from that calculated using
the phenomenological potential 1. Furthermore, the double folding potential
used has the same volume integrals as the phenomenological potential 1. In
Ref. \cite{ak97} we left open the choice of the potential we use to extract
the value of the ANC for the system $^{9}$Be+p$\rightarrow ^{10}$B, and two
slightly different values were extracted using potentials 1 and 2. The
present study and the calculations made with the double folding potential
indicate that we can select potential 1 as the only potential, and that the
value C$_{1}^{2}$=4.91(19) fm$^{-1}$ is a better choice than the weighted
average C$^{2}$=5.06(46) fm$^{-1}$ given previously. For the reaction $^{13}$%
C($^{14}$N,$^{13}$C)$^{14}$N we found that the value of the cross section
calculated using the double folding potential JLM(1) varies at any angle
between $\theta _{cm}=0^{\circ }-35^{\circ }$ by less than 2\% from that
calculated using potential 1 in Table \ref{tab2}, and its integral over the
same angular range does not vary at all. This is easy to understand given
the fact that the surface part of the nuclear potential is the contributing
factor in the description of both the elastic scattering and the transfer
reaction. Previously we found that the calculated cross sections for the
proton transfer reaction $^{13}$C($^{14}$N,$^{13}$C)$^{14}$N differ by about
2\% between any consecutive phenomenological potential families in Table \ref
{tab2}, as described in Ref. \cite{tra98}. The present verification
increases our confidence in using the double folding procedure for the
description of the transfer reactions.

\section{Optical potentials for $^{10}$B($^{7}$Be,$^{8}$B)$^{9}$Be and $%
^{14} $N($^{7}$Be,$^{8}$B)$^{13}$C reactions}

\label{sec6}Using the procedure outlined above, the JLM(1) effective
nucleon-nucleon interaction and the average renormalization coefficients
extracted, we calculated the optical model potentials needed in the analysis
of the proton transfer reactions $^{10}$B($^{7}$Be,$^{8}$B)$^{9}$Be and $%
^{14}$N($^{7}$Be,$^{8}$B)$^{13}$C at E($^{7}$Be)=84 MeV, which were the
original motivation for the present study. The systems involve the
radioactive $^{7}$Be and $^{8}$B nuclei, both loosely bound and with
important clusterization in their ground states. This made us treat the
Hartree-Fock densities carefully, forcing the calculations to reproduce the
correct binding energies through slight modifications in the surface term as
stated before. Furthermore, in the final calculations we imposed
self-consistency by requiring that the tail of the density distribution of $%
^{8}$B have the asymptotic behavior given by the ANC extracted from the
transfer reaction (but not predicted by HF). This produced changes in the
potentials only at distances larger than 8 fm, as shown in Figure \ref{fig8}%
, and did not introduce any substantial change in the calculated proton
transfer cross sections. The optical model potentials obtained reproduced
very well the measured angular distributions for the elastic scattering of $%
^{7}$Be on the $^{10}$B (Ref. \cite{aa99}) and $^{14}$N (Ref. \cite{aa99b})
targets without any need for further adjustments (in both cases elastic
scattering was actually calculated using JLM(1) potentials not only on the
main component of the target, $^{10}$B and $^{14}$N, respectively, but also
on the $^{16}$O and $^{12}$C nuclei present as impurities in the $^{10}$B
and in the melamine target, respectively). This provided confidence for
using the extracted potentials for the description of the angular
distributions of the proton transfer reactions. Again, the shape of the
measured angular distributions are very well reproduced, as seen for $^{14}$%
N($^{7}$Be,$^{8}$B)$^{13}$C in Figure \ref{fig9}. In turn, the calculated
cross sections were used to extract the Asymptotic Normalization Coefficient
for the system $^{8}$B$\rightarrow ^{7}$Be+p and, consequently, the
astrophysical factor S$_{17}(0)$ reported in Refs.\cite{aa99,aa99b}. The
potentials obtained for these nucleus-nucleus combinations involving
radioactive $^{7}$Be and $^{8}$B are not exactly of Woods-Saxon shape, but
can be approximated in the region of their surface by Woods-Saxon
potentials. In Table \ref{tab5} we give the parameters found by fitting the
range of radii r=2-12 fm.

In order to estimate the uncertainty in the ANCs due to the optical model
potentials, we consider that the standard deviations of the normalization
coefficients $\delta $N$_{V}$=0.014 and $\delta $N$_{W}$=0.087 give a good
measure of the uncertainty with which we can find the depths of the real and
imaginary potentials wells, respectively. By the choice of the systems
considered here, we span a good range of p-shell nuclei, averaging
properties similar to those of radioactive ones in terms of mass, separation
energy, structure of the ground states, incident energies, number of open
reaction channels, etc. We used these standard deviations around the average
value of the renormalization coefficients to evaluate the uncertainty in
extracting the ANCs. The uncertainties arise through the DWBA calculations
of the transfer reaction cross section. We took the geometry as given by the
double folding procedure and determined the variation of the calculated
proton transfer cross section integrated over the angular range relevant in
the experiments. The potential depths were varied from N$_{V}$-$\delta $N$%
_{V}$ to N$_{V}$+$\delta $N$_{V}$ for the real part and from N$_{W}$-$\delta 
$N$_{W}$ to N$_{W}$+$\delta $N$_{W}$ for the imaginary part for the entrance
and exit channels independently and the resulting variations were added in
quadrature to estimate the relative uncertainty in the DWBA calculations.
With this procedure we found a 7.5\% uncertainty in the calculated $^{10}$B($%
^{7}$Be,$^{8}$B)$^{9}$Be transfer cross section due to DWBA calculations.
The same procedure gave an estimate of 7.7\% for the uncertainty due to DWBA
calculations for the $^{14}$N($^{7}$Be,$^{8}$B)$^{13}$C reaction. Most of
the contribution comes from the uncertainty in the imaginary renormalization
coefficient (7.5\%), while the real one contributes only about a quarter of
that (2\%). Note that in varying separately the depths of the potentials in
the entrance and exit channels for the same reaction, we treat the
uncertainties as uncorrelated between the channels involving $^{7}$Be and $%
^{8}$B, respectively, whereas the uncertainties between the two different
reactions remain correlated through the use of the same procedure and of the
same average values for the renormalization coefficients. When we treated
the uncertainties in assessing the depths of the potentials in the entrance
and exit channels as totally correlated, as we did in Ref. \cite{aa99}, we
obtained an uncertainty of about 10\%.

We note here that given the observed strong dependence of the calculations
on the imaginary part of the potential, and the relative independence of the
real and imaginary parts of the potential, the measurement of the total
reaction cross section of $^{7}$Be and $^{8}$B might be useful. It can set
an extra constraint on the renormalization of the imaginary part of the
potential, and eventually decrease the uncertainty in the potential used,
and therefore in the DWBA results.

\section{Conclusions}

\label{concl} We have measured the elastic scattering of $^{7}$Li, $^{10}${%
B, }$^{13}${C and }$^{14}${N} on $^{9}${Be and }$^{13}${C targets} at or
around E/A=10 MeV/nucleon for angular ranges up to around the nuclear
rainbow angle, using a fine angle binning of $\Delta \theta
_{lab}=0.5^{\circ }$. All these projectile-target combinations, with the
exception of $^{14}$N+$^{13}$C, have in common the fact that one or both
partners are weakly bound and we expect contributions from break-up channels
to be important. At the same time it is known that at energies around 10
MeV/nucleon the contribution of the dynamic polarization potential is
non-negligible. Parameters for optical model potentials of Woods-Saxon form
were obtained from the fit of the elastic scattering angular distributions.
In addition, nucleus-nucleus potentials were calculated by a double folding
procedure using six different effective nucleon-nucleon interactions. The
nuclear densities calculated for each partner in the Hartree-Fock
approximation were folded with four different versions of the M3Y
nucleon-nucleon interaction and with the effective interaction of Jeukenne
et al. \cite{jlm}, calculated with two different techniques to account for
the local density. The resulting nucleus-nucleus potentials were later
renormalized to obtain a fit of the elastic scattering data. The
normalization constants have similar values in all systems for each
effective interaction used, which makes it appear likely that the procedure
can be extended to the calculation of optical potentials for other similar
nucleus-nucleus systems. From all effective interactions used, we conclude
that JLM(1) gives the best results. It provides us with an imaginary part
that has a geometry which is independent from that of the real part of the
potential. The imaginary well produced is wider than the real one, as the
fit of the data with phenomenological Woods-Saxon wells requires. At the
same time, it gives the least spread in the value of the renormalization
coefficients, which suggests that its density dependence accounts very well
for the differences between the nuclei involved, particularly in the surface
region. We find that while the depth of the real potential needs a
substantial renormalization ($\langle $N$_{V}\rangle $=0.366$\pm $0.014),
the imaginary part needs no such renormalization ($\langle $N$_{W}\rangle $%
=1.000$\pm $0.087). This also suggests that the imaginary part of the
effective interaction is well accounted for. The need for a substantial
renormalization of the real part was attributed to the effect of the
break-up channels, which are very important in nuclei with low binding
energies like those encountered in our experiments \cite{satch}. This
suggests that the average value of the renormalization constant for the real
potential depth found above is valid for the region of p-shell nuclei
considered in this study and might be somewhat different in other regions.

The renormalized double folded potentials obtained were also used in the
DWBA analysis of the proton transfer reactions with stable nuclei, and the
results were found to be in excellent agreement with those given by the
phenomenological Woods-Saxon potentials. We found that comparison of the
double folded potentials and the phenomenological ones gives a way to select
between the different families based on their volume integrals.

Finally, the procedure found was applied to extract the optical model
potentials for the $^{7}$Be and $^{8}$B radioactive projectiles needed in
the description of the $^{7}$Be+$^{10}$B and $^{7}$Be+$^{14}$N experiments.
Good description for the elastic scattering data is found without any need
for readjustment of the shape or magnitude of the angular distributions. The
shape of the angular distributions measured for the proton transfer
reactions $^{10}$B($^{7}$Be,$^{8}$B)$^{9}$Be and $^{14}$N($^{7}$Be,$^{8}$B)$%
^{13}$C are also very well predicted. The calculated DWBA cross sections
were used to extract the ANC for the $^{8}$B$\rightarrow ^{7}$Be+p system
from each reaction, and consequently the astrophysical S$_{17}$(0) factor
reported in Refs. \cite{aa99,aa99b}. Furthermore, we used the standard
deviations of the renormalization coefficients $\delta $N$_{V}$, $\delta $N$%
_{W}$ to estimate the contribution of the DWBA calculations to the
uncertainty of the determined ANC and thus S$_{17}$(0). We found this
contribution to be around 8\%.

\acknowledgements
One of us (FC) acknowledges the support of the Cyclotron Institute, Texas
A\&M University, for a stay in College Station, TX, during which a part of
this work was done. This work was supported in part by the U. S. Department
of Energy under Grant no DE-FG03-93ER40773 and by the Robert A. Welch
Foundation.

\begin{figure}[tbp]
\caption{Spectrum from the elastic scattering of $^{7}${Li} on the $^{9}${Be}
target at E$_{lab}=130$ MeV and $\theta_{lab}=27.25^{\circ} \pm 0.25^{\circ}$%
. The peak labeled D is a combination of inelastic excitation of $^{9}$Be
(2.9 MeV) and of double excitation of the target and projectile. }
\label{fig1}
\end{figure}

\begin{figure}[tbp]
\caption{Angular distributions for the elastic scattering of a) $^{10}$B
(100 MeV) + $^9$Be, b) $^{13}$C (130 MeV)+ $^9$Be, and c) $^{14}$N (162 MeV)+%
$^{13}$C. The curves are fits with the potentials presented in Table II.}
\label{fig2}
\end{figure}

\begin{figure}[tbp]
\caption{Same as Figure 2, for the systems a) $^7$Li(63 MeV)+$^9$Be, b) $^7$%
Li(63 MeV)+$^{13}$C, c) $^7$Li(130 MeV)+$^9$Be, and d) $^7$Li(130 MeV)+$%
^{13} $C. }
\label{fig3}
\end{figure}

\begin{figure}[tbp]
\caption{ The near-side/far-side decomposition of the elastic scattering for
a) $^{10}$B+$^{9}$Be, b) $^{14}$N+$^{13}$C, and c) $^7$Li (63 MeV) + $^{13}$%
C. }
\label{fig4}
\end{figure}

\begin{figure}[tbp]
\caption{ Fit of the angular distributions with the folding potentials of
Table IV. The curves are labeled: M3Y/ZR for the M3Y zero-range interaction,
BDM3Yn for the density-dependent M3Y interactions, and JLM(n) for the
interaction of Jeukenne, Lejeune and Mahaux, respectively. The cases
presented are (a) $^{14}$N+$^{13}$C, (b) $^{10}$B+$^9$Be, (c) $^7$Li(63 MeV)+%
$^{13}$C, and (d) $^7$Li(63 MeV)+$^9$Be. }
\label{fig5}
\end{figure}

\begin{figure}[tbp]
\caption{ Same as Figure \ref{fig5}, but for the systems: (a) $^7$Li(130
MeV)+$^{13}$C, (b) $^{13}$C+$^9$Be, (c) $^6$Li(99 MeV)+$^{12}$C and (d) $^6$%
Li(99 MeV)+$^{28}$Si (data from \protect\cite{schw}).}
\label{fig6}
\end{figure}

\begin{figure}[tbp]
\caption{The renormalization coefficients extracted for the double folding
potentials calculated with the six effective nucleon-nucleon interactions,
as described in the text. The projectile-target combinations are those of
Table \ref{tab4}.}
\label{fig7}
\end{figure}

\begin{figure}[tbp]
\caption{The double folded potentials calculated with the standard
Hartree-Fock mass distributions (dashed lines) are compared with those
obtained when the tail of the proton distribution of $^8$B is given by the
ANC obtained from our experiments (full line). Both real (V) and imaginary
(W) potentials are shown for the system $^8$B + $^9$Be, using the JLM(1)
effective interaction. }
\label{fig8}
\end{figure}

\begin{figure}[tbp]
\caption{ The angular distribution for the elastic proton transfer $^{14}$N($%
^7$Be,$^8$B)$^{13}$C at E$_{lab}$=84 MeV, calculated using the optical model
potential obtained with the JLM(1) effective interaction (dashed line) is
compared with the one smoothed by a Monte Carlo procedure to account for the
experimental resolution (solid line) and with the experimental points. The
dotted and dash-dotted lines represent the calculated cross section (not
smoothed) with the imaginary potential depths renormalized by N$_W\pm \delta$%
N$_W$ (upper panel). The lower panel presents the ratios of the transfer
cross sections calculated using renormalization coefficients for the
imaginary part of the potential N$_W+\delta$N$_W$ (dotted line), N$_W$
(solid line) and N$_W-\delta$N$_W$ (dash-dotted line) to that calculated
with the median value N$_W$.}
\label{fig9}
\end{figure}

\newpage

\tightenlines

\begin{center}
\begin{table}[tbp]
\caption{ List of the elastic scattering experiments presented in this
paper. }
\label{tab1}
\end{table}
\begin{tabular}{||c|c|c|c||}
\hline
&  &  &  \\ 
No. & Projectile-target & E (MeV) & $\theta _{lab}(\deg .)$ \\ 
&  &  &  \\ \hline
&  &  &  \\ 
1 & $^{10}$B + $^{9}$Be & 100 & 4 - 30 \\ 
2 & $^{13}$C + $^{9}$Be & 130 & 4 - 22 \\ 
3 & $^{14}$N + $^{13}$C & 162 & 2 - 34 \\ 
4 & $^{7}$Li + $^{9}$Be & 63 & 4 - 52 \\ 
5 & $^{7}$Li + $^{13}$C & 63 & 4 - 56 \\ 
6 & $^{7}$Li + $^{9}$Be & 130 & 4 - 47 \\ 
7 & $^{7}$Li + $^{13}$C & 130 & 4 - 47 \\ 
&  &  &  \\ \hline
\end{tabular}
\end{center}

\newpage

\begin{table}[tbp]
\caption{ The parameters of the Woods-Saxon optical model potentials
extracted from the analysis of the elastic scattering data for
projectile-target combinations studied here. r$_{C}$=1 fm for all
potentials. }
\label{tab2}
\end{table}
\vskip.7cm {{\footnotesize 
\begin{tabular}{||c|c|c|c|c|c|c|c|c|c|c|c|c|c||}
\hline
&  &  &  &  &  &  &  &  &  &  &  &  &  \\ 
Channel & Pot. & V & W & r$_{V}$ & r$_{W}$ & a$_{V}$ & a$_{W}$ & $\chi ^{2}$
& $\sigma _{R}$ & J$_{V}$ & R$_{V}$ & J$_{W}$ & R$_{W}$ \\ 
&  & [MeV] & [MeV] & [fm] & [fm] & [fm] & [fm] &  &  & [MeV fm$^{3}$] & [fm]
& [MeV fm$^{3}$] & [fm] \\ 
&  &  &  &  &  &  &  &  &  &  &  &  &  \\ \hline
&  &  &  &  &  &  &  &  &  &  &  &  &  \\ 
$^{10}$B (100 MeV) + $^{9}$Be & 1 & 64.2 & 30.1 & 0.78 & 0.99 & 0.99 & 0.75
& 19.8 & 1318 & 206 & 4.51 & 136 & 4.28 \\ 
& 2 & 131.2 & 29.7 & 0.67 & 0.95 & 0.90 & 0.86 & 45.4 & 1411 & 276 & 3.99 & 
131 & 4.46 \\ 
& 3 & 203.2 & 24.7 & 0.81 & 1.04 & 0.60 & 0.83 & 61.8 & 1428 & 499 & 3.46 & 
133 & 4.59 \\ 
&  &  &  &  &  &  &  &  &  &  &  &  &  \\ \hline
&  &  &  &  &  &  &  &  &  &  &  &  &  \\ 
$^{14}${N} (162 MeV) + $^{13}${C} & 1 & 79.22 & 30.27 & 0.96 & 1.05 & 0.76 & 
0.72 & 17.4 & 1542 & 221 & 4.52 & 105 & 4.69 \\ 
& 2 & 134.76 & 35.23 & 0.88 & 1.05 & 0.75 & 0.67 & 18.3 & 1525 & 299 & 4.28
& 120 & 4.61 \\ 
& 3 & 176.03 & 35.84 & 0.86 & 1.07 & 0.72 & 0.65 & 23.3 & 1527 & 361 & 4.15
& 125 & 4.62 \\ 
& 4 & 241.36 & 37.45 & 0.82 & 1.06 & 0.71 & 0.66 & 27.5 & 1533 & 438 & 4.00
& 129 & 4.61 \\ 
& 5 & 306.44 & 39.14 & 0.81 & 1.05 & 0.68 & 0.68 & 36.1 & 1552 & 522 & 3.90
& 132 & 4.61 \\ 
&  &  &  &  &  &  &  &  &  &  &  &  &  \\ \hline
&  &  &  &  &  &  &  &  &  &  &  &  &  \\ 
$^{13}${C} (130 MeV) + $^{9}${Be} & 1 & 94.2 & 20.9 & 0.77 & 0.99 & 0.87 & 
0.97 & 15.0 & 1592 & 223 & 4.19 & 94 & 4.96 \\ 
& 2 & 164.2 & 23.0 & 0.67 & 0.98 & 0.86 & 0.95 & 14.2 & 1576 & 283 & 3.94 & 
99 & 4.87 \\ 
& 3 & 226.7 & 31.9 & 0.62 & 0.90 & 0.85 & 0.95 & 14.8 & 1573 & 328 & 3.81 & 
113 & 4.70 \\ 
&  &  &  &  &  &  &  &  &  &  &  &  &  \\ \hline
&  &  &  &  &  &  &  &  &  &  &  &  &  \\ 
$^{7}${Li} (63 MeV) + $^{9}${Be} & 1 & 134.4 & 19.82 & 0.54 & 1.03 & 0.95 & 
0.92 & 8.0 & 1414 & 267 & 3.90 & 137 & 4.66 \\ 
& 2 & 221.6 & 27.33 & 0.54 & 0.92 & 0.83 & 0.97 & 10.6 & 1449 & 367 & 3.50 & 
153 & 4.60 \\ 
& 3 & 276.5 & 37.3 & 0.61 & 0.81 & 0.72 & 1.02 & 15.7 & 1482 & 499 & 3.27 & 
158 & 4.64 \\ 
&  &  &  &  &  &  &  &  &  &  &  &  &  \\ \hline
$^{7}${Li} (63 MeV) + $^{13}${C} & 1 & 54.3 & 29.9 & 0.92 & 1.03 & 0.79 & 
0.69 & 28.8 & 1318 & 209 & 4.21 & 144 & 4.26 \\ 
& 2 & 99.8 & 22.0 & 0.77 & 1.01 & 0.81 & 0.81 & 21.6 & 1363 & 257 & 3.92 & 
109 & 4.49 \\ 
& 3 & 154.8 & 22.7 & 0.76 & 1.00 & 0.71 & 0.83 & 19.8 & 1378 & 357 & 3.64 & 
111 & 4.51 \\ 
& 4 & 244.6 & 26.4 & 0.68 & 0.96 & 0.71 & 0.84 & 20.4 & 1382 & 438 & 3.47 & 
117 & 4.45 \\ 
&  &  &  &  &  &  &  &  &  &  &  &  &  \\ \hline
$^{7}${Li} (130 MeV) + $^{9}${Be} & 1 & 60.0 & 17.71 & 0.86 & 1.07 & 0.65 & 
1.12 & 150. & 1564 & 217 & 3.58 & 154 & 5.33 \\ 
& 2 & 129.4 & 30.7 & 0.57 & 0.80 & 0.90 & 1.17 & 208 & 1488 & 261 & 3.77 & 
158 & 5.02 \\ 
&  &  &  &  &  &  &  &  &  &  &  &  &  \\ \hline
$^{7}${Li} (130 MeV) + $^{13}${C} & 1 & 123.3 & 32.74 & 0.76 & 0.94 & 0.76 & 
0.90 & 79.1 & 1406 & 297 & 3.79 & 145 & 4.66 \\ 
& 2 & 157.9 & 31.97 & 0.63 & 0.90 & 0.87 & 0.94 & 77.3 & 1393 & 289 & 3.83 & 
133 & 4.59 \\ 
& 3 & 201.9 & 25.59 & 0.73 & 1.03 & 0.69 & 0.86 & 129. & 1418 & 419 & 3.52 & 
142 & 4.66 \\ 
& 4 & 300.0 & 30.78 & 0.73 & 0.98 & 0.64 & 0.89 & 147. & 1441 & 543 & 3.37 & 
150 & 4.63 \\ 
&  &  &  &  &  &  &  &  &  &  &  &  &  \\ \hline
\end{tabular}
}} \newpage

\begin{center}
\begin{table}[tbp]
\caption{ Radii and binding energies of the calculated Hartree-Fock one-body
densities, compared with the experimental data. R$_{p}$, R$_{n}$, R$_{m}$
and R$_{ch}$ stand for the root mean square radii of the calculated proton,
neutron, mass and charge distributions respectively and R$_{ch}^{exp}$ is
the experimental charge rms. B are the binding energies.}
\label{tab_dens}
\end{table}
{\footnotesize \vskip.7cm 
\begin{tabular}{||c|c|c|c|c|c|c|c||}
\hline
&  &  &  &  &  &  &  \\ 
nucleus & R$_{p}$ & R$_{n}$ & R$_{m}$ & R$_{ch}$ & R$_{ch}^{exp}$ & B$_{th}$
& B$_{exp}$ \\ 
& [fm] & [fm] & [fm] & [fm] & [fm] & [MeV] & [MeV] \\ 
&  &  &  &  & Ref. \cite{Vri87} &  & Ref. \cite{AWa93} \\ 
&  &  &  &  &  &  &  \\ \hline
&  &  &  &  &  &  &  \\ 
$^{6}${Li} & 2.21 & 2.20 & 2.20 & 2.21 & 2.54(5) & 31.929 & 31.994 \\ 
$^{7}${Li} & 2.15 & 2.35 & 2.26 & 2.16 & 2.39(3) & 39.234 & 39.244 \\ 
$^{7}${Be} & 2.37 & 2.14 & 2.28 & 2.38 & 2.36(2)$^{*}$ & 37.606 & 37.600 \\ 
$^{8}${B} & 2.57 & 2.18 & 2.43 & 2.58 & 2.45(5)$^{*}$ & 37.744 & 37.737 \\ 
$^{9}${Be} & 2.26 & 2.39 & 2.33 & 2.29 & 2.50(9) & 58.203 & 58.164 \\ 
$^{10}${B} & 2.40 & 2.39 & 2.40 & 2.45 & 2.45(12) & 64.631 & 64.750 \\ 
$^{12}${C} & 2.44 & 2.43 & 2.44 & 2.49 & 2.47(2) & 92.149 & 92.161 \\ 
$^{13}${C} & 2.47 & 2.56 & 2.52 & 2.53 & 2.440(25) & 97.135 & 97.108 \\ 
$^{14}${N} & 2.58 & 2.57 & 2.57 & 2.64 & 2.58(2) & 104.246 & 104.658 \\ 
&  &  &  &  &  &  &  \\ \hline
\end{tabular}
}
\end{center}

$\ast$ proton density $rms$ radius obtained by Tanihata et al \cite{tanih}
from interaction cross sections.

\newpage \tightenlines

\begin{center}
\begin{table}[tbp]
\caption{ Best fit renormalisation parameters N$_{V}$ and N$_{W}$ for
folding potentials with various effective interactions (see eqs. (15) and
(16)). For each reaction channel, the values of N$_{V}$ are given in the
first line and N$_{W}$ in the second line. For each effective interaction,
the mean values and dispersions are given in the last two lines. Only cases
2-7 are used to determine averages, as described in the text.}
\label{tab4}
\end{table}
\vskip.7cm 
\begin{tabular}{||c|c|c|c|c|c|c|c||}
\hline
&  &  &  &  &  &  &  \\ 
No. & Projectile-target & JLM(1) & JLM(2) & M3Y/ZR & M3Y/FR & BDM3Y1 & BDM3Y3
\\ 
&  &  &  &  &  &  &  \\ \hline
&  &  &  &  &  &  &  \\ 
1 & $^{14}${N} (162 MeV) + $^{13}${C} & 0.456 & 0.509 & 0.778 & 1.275 & 0.721
& 0.832 \\ 
&  & 0.844 & 0.996 & 0.469 & 0.887 & 0.419 & 0.492 \\ \hline
2 & $^{10}${B} (100 MeV) + $^{9}${Be} & 0.368 & 0.387 & 0.516 & 0.667 & 0.584
& 0.668 \\ 
&  & 1.168 & 1.131 & 0.571 & 1.116 & 0.506 & 0.596 \\ \hline
3 & $^{13}${C} (130 MeV) + $^{9}${Be} & 0.369 & 0.413 & 0.489 &  & 0.576 & 
0.648 \\ 
&  & 0.937 & 1.124 & 0.726 &  & 0.550 & 0.656 \\ \hline
4 & $^{7}${Li} (63 MeV) + $^{13}${C} & 0.323 & 0.364 & 0.588 & 0.787 & 0.552
& 0.634 \\ 
&  & 1.00 & 1.007 & 0.503 & 0.831 & 0.458 & 0.535 \\ \hline
5 & $^{7}${Li} (63 MeV) + $^{9}${Be} & 0.360 & 0.403 & 0.588 & 0.759 & 0.568
& 0.645 \\ 
&  & 1.00 & 1.438 & 0.818 & 1.175 & 0.733 & 0.864 \\ \hline
6 & $^{7}${Li} (130 MeV) + $^{13}${C} & 0.380 & 0.418 & 0.595 & 0.914 & 0.571
& 0.651 \\ 
&  & 0.957 & 1.077 & 0.508 & 0.893 & 0.472 & 0.547 \\ \hline
7 & $^{7}${Li} (130 MeV) + $^{9}${Be} & 0.368 & 0.413 & 0.489 & 0.806 & 0.576
& 0.648 \\ 
&  & 0.937 & 1.124 & 0.726 & 1.110 & 0.550 & 0.656 \\ \hline\hline
8 & $^{6}${Li} (99 MeV) + $^{12}${C} & 0.449 & 0.493 & 0.716 & 1.178 & 0.687
& 0.785 \\ 
&  & 1.044 & 1.166 & 0.536 & 0.942 & 0.510 & 0.585 \\ \hline
9 & $^{6}${Li} (99 MeV) + $^{28}${Si} & 0.368 & 0.408 & 0.565 & 0.960 & 0.534
& 0.611 \\ 
&  & 1.168 & 1.324 & 0.683 & 1.170 & 0.621 & 0.726 \\ \hline
10 & $^{7}${Li} (63 MeV) + $^{12}${C} & 0.278 & 0.309 & 0.502 &  & 0.478 & 
0.546 \\ 
&  & 0.746 & 0.920 & 0.464 &  & 0.423 & 0.493 \\ \hline
11 & $^{7}${Li} (79 MeV) + $^{12}${C} & 0.315 & 0.347 & 0.521 &  & 0.505 & 
0.573 \\ 
&  & 0.864 & 1.009 & 0.458 &  & 0.426 & 0.493 \\ \hline\hline
& average of cases 2-7 & 0.366$\pm $0.014 & 0.405$\pm $0.017 & 0.553$\pm $%
0.062 & 0.787$\pm $0.089 & 0.578$\pm $0.010 & 0.658$\pm $0.013 \\ 
&  & 1.000$\pm $0.087 & 1.143$\pm $0.145 & 0.631$\pm $0.131 & 1.025$\pm $%
0.153 & 0.553$\pm $0.082 & 0.631$\pm $0.115 \\ \hline\hline
\end{tabular}

\begin{table}[tbp]
\caption{ Parameters of volume Woods-Saxon type potentials that best fit the
nuclear part of the numerical potentials obtained with the double folding
procedure using the JLM(1) effective interaction in the range r=2-12 fm (see
text). Renormalization of the depths is included. R$_V$ and R$_W$ are the
half-radii of the potentials.}
\label{tab5}
\end{table}
\vskip.7cm 
\begin{tabular}{||c|c|c|c|c|c|c|c|c|c||}
\hline
Projectile-target & E$_{inc}$ & V & W & R$_V$ & R$_W$ & a$_V$ & a$_W$ & J$_V$
& J$_W$ \\ 
& [MeV] & [MeV] & [MeV] & [fm] & [fm] & [fm] & [fm] & [MeV fm$^3$] & [MeV fm$%
^3$] \\ 
&  &  &  &  &  &  &  &  &  \\ \hline
$^7$Be + $^{10}$B & 84 & 63.8 & 29.4 & 3.18 & 3.49 & 0.85 & 0.95 & 210 & 130
\\ 
$^8$B + $^9$Be & 81 & 67.0 & 31.8 & 3.18 & 3.54 & 0.88 & 0.99 & 236 & 145 \\ 
$^7$Be + $^{14}$N & 84 & 79.1 & 36.0 & 3.30 & 3.62 & 0.88 & 0.98 & 207 & 126
\\ 
$^8$B + $^{13}$C & 78 & 85.2 & 39.3 & 3.30 & 3.76 & 0.91 & 1.02 & 216 & 145
\\ \hline
\end{tabular}
\end{center}

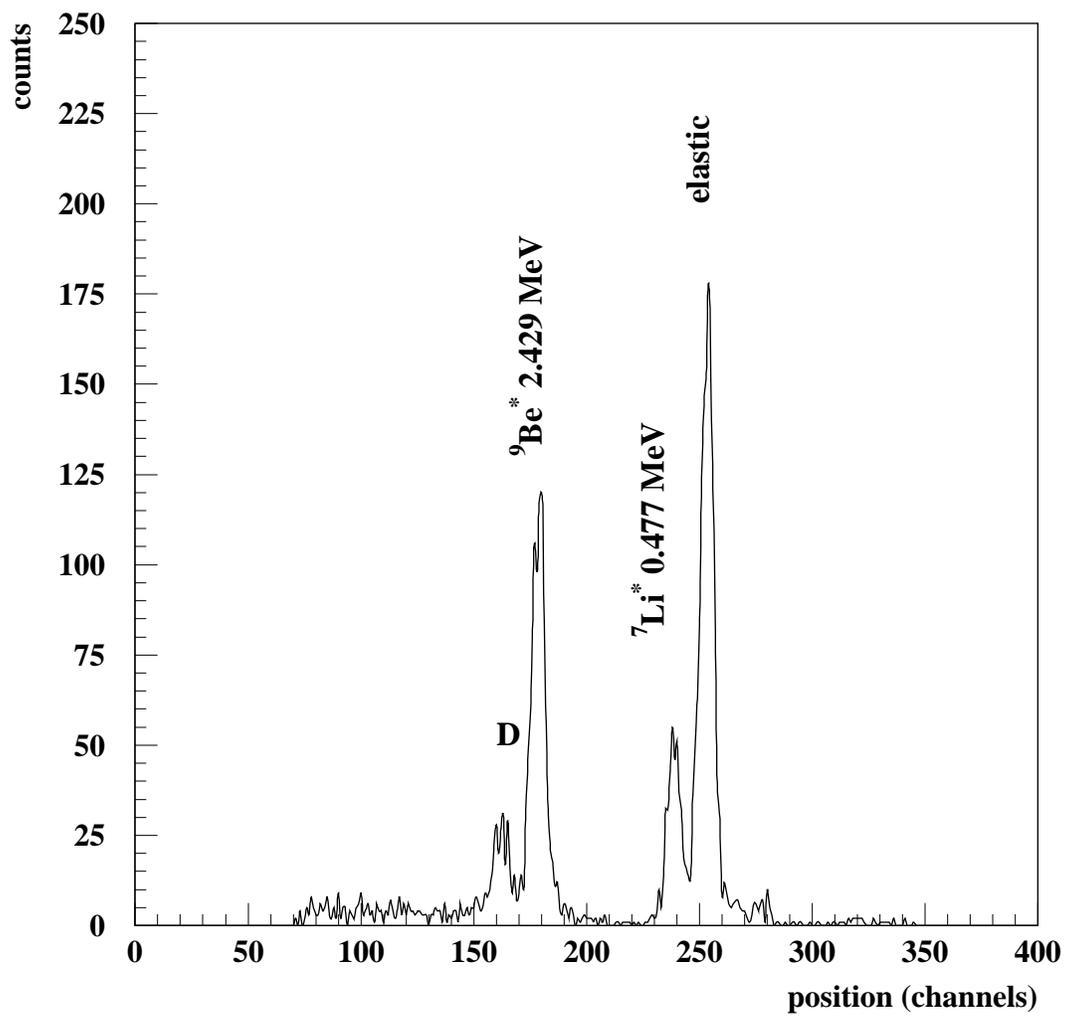
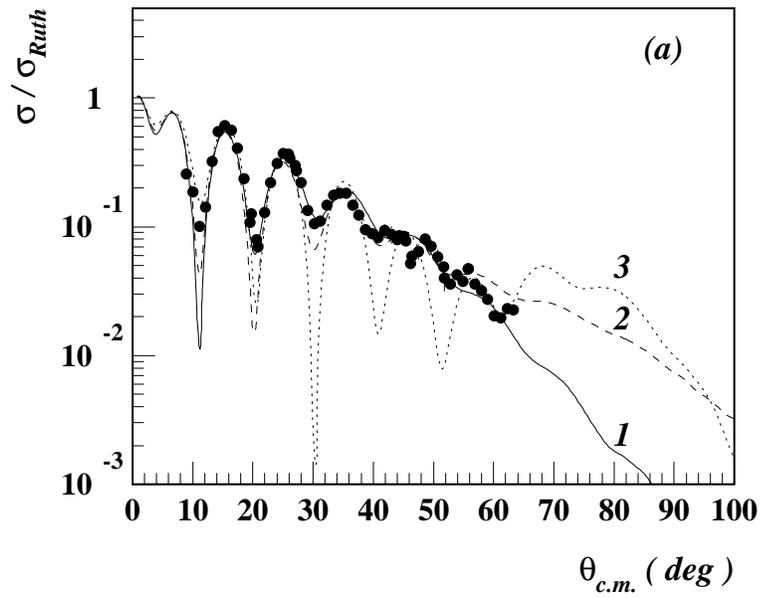
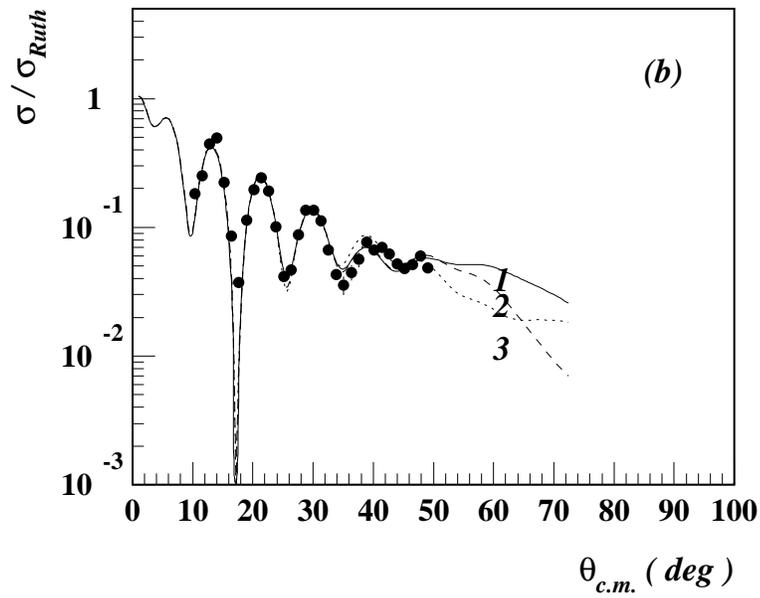
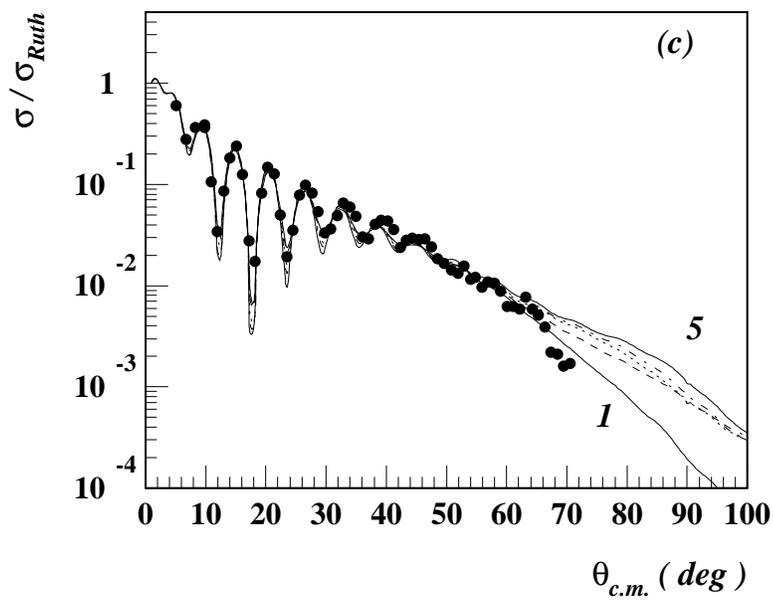
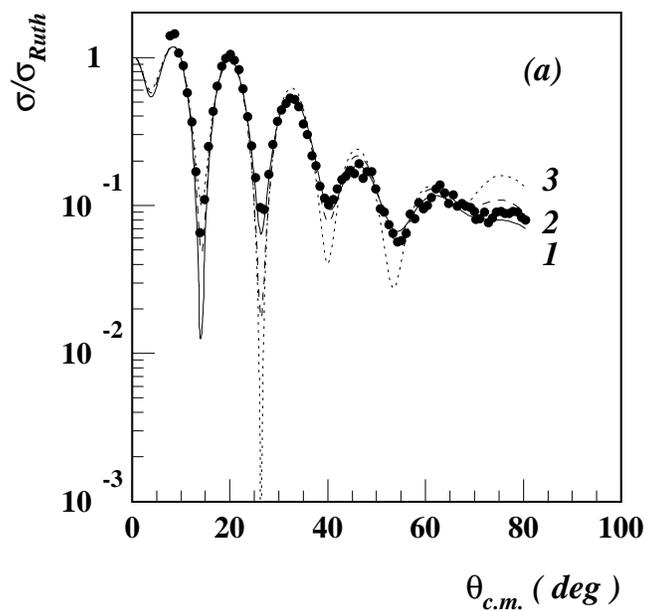
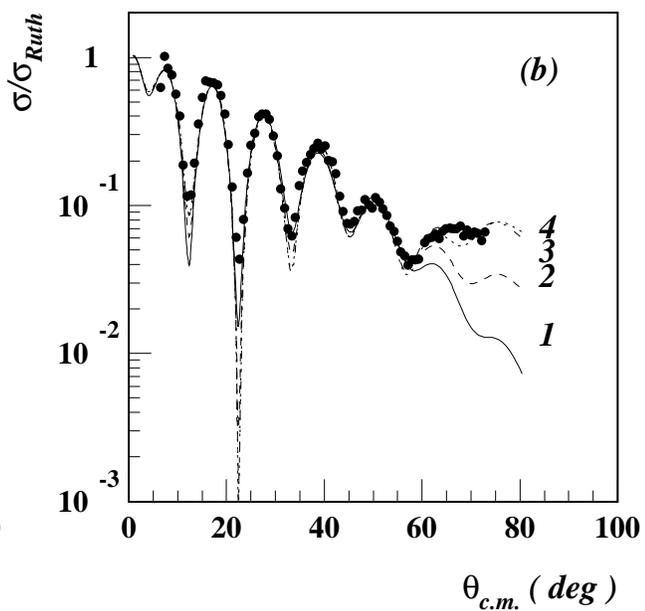
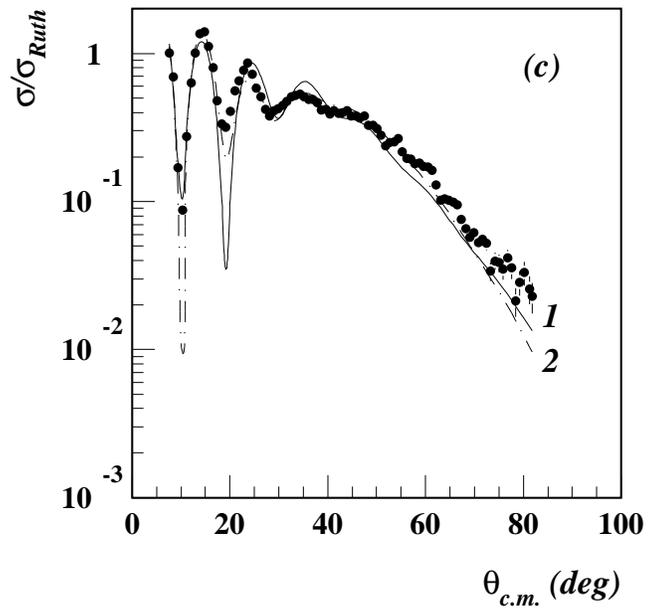
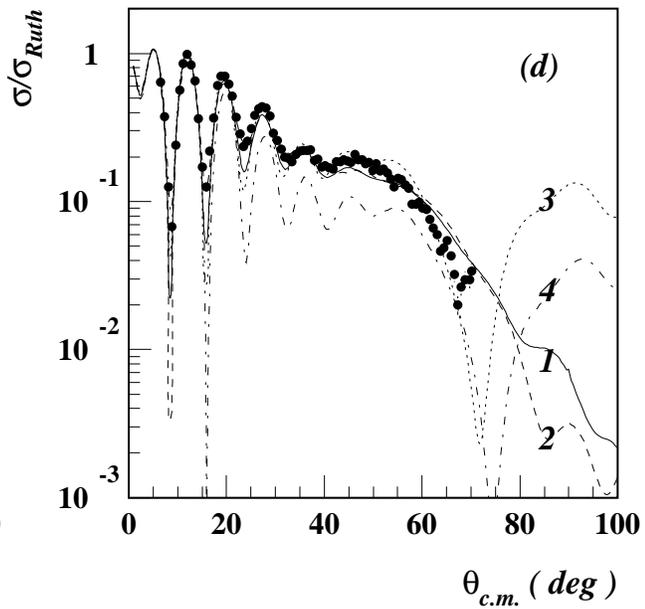
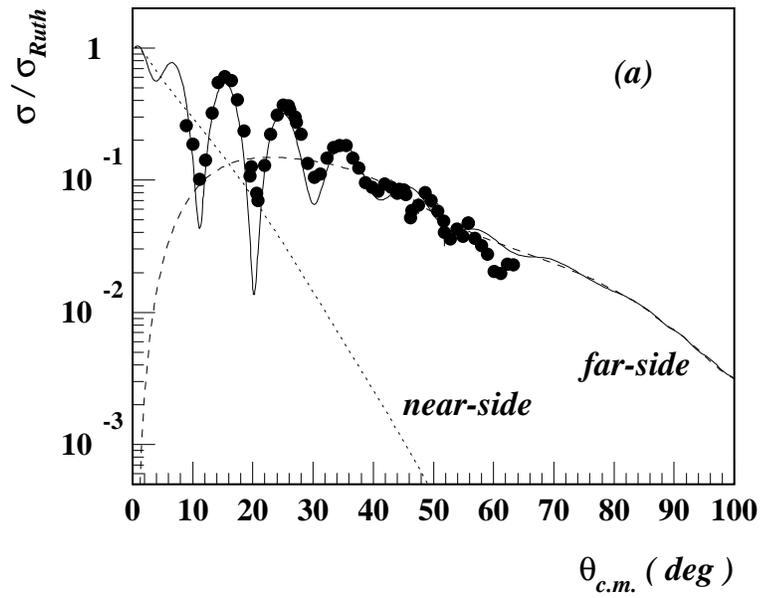
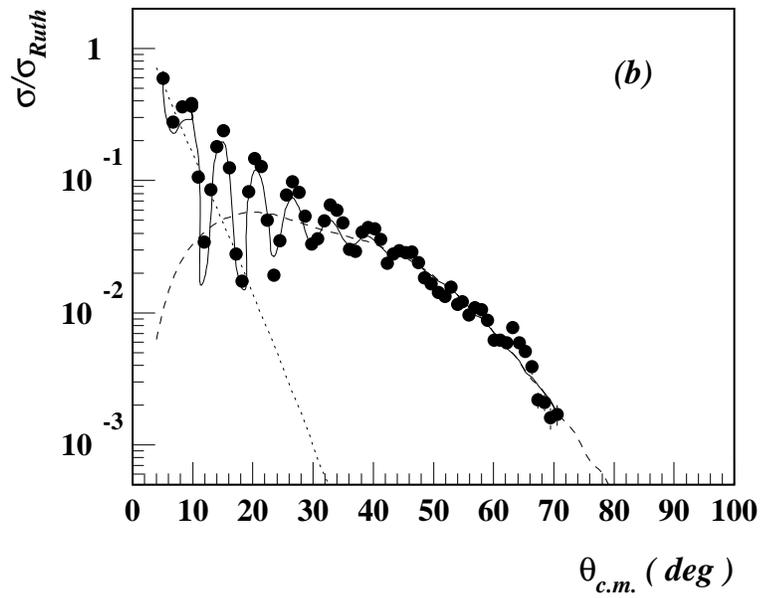
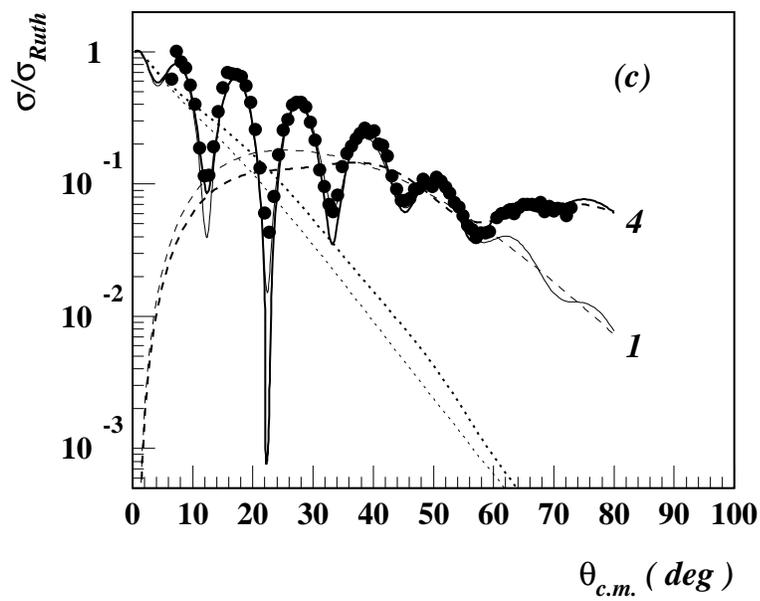
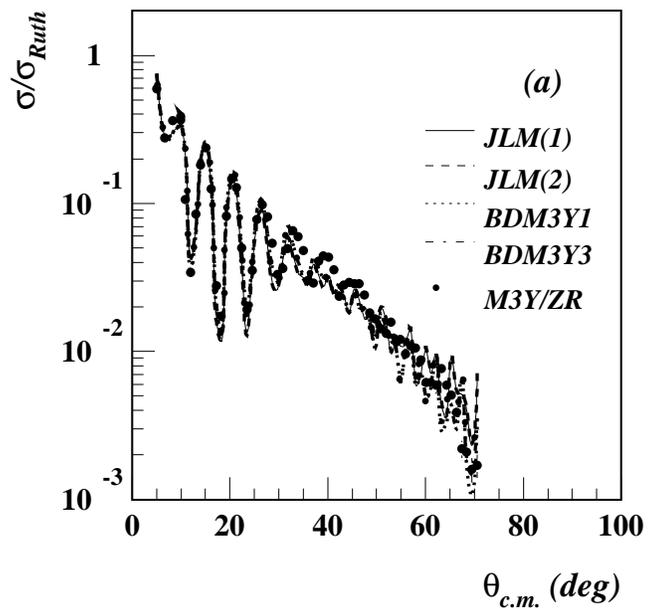
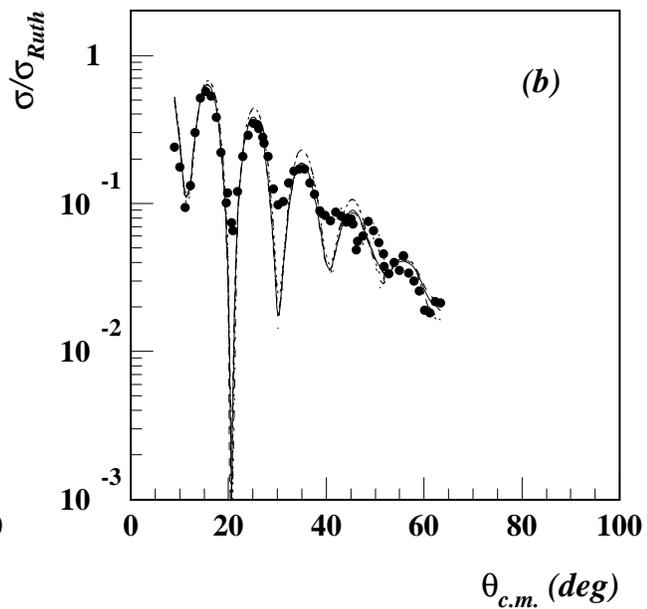
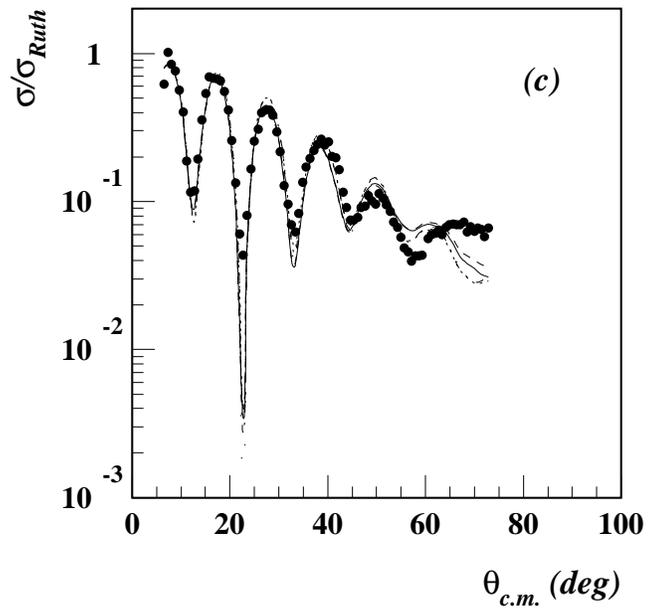
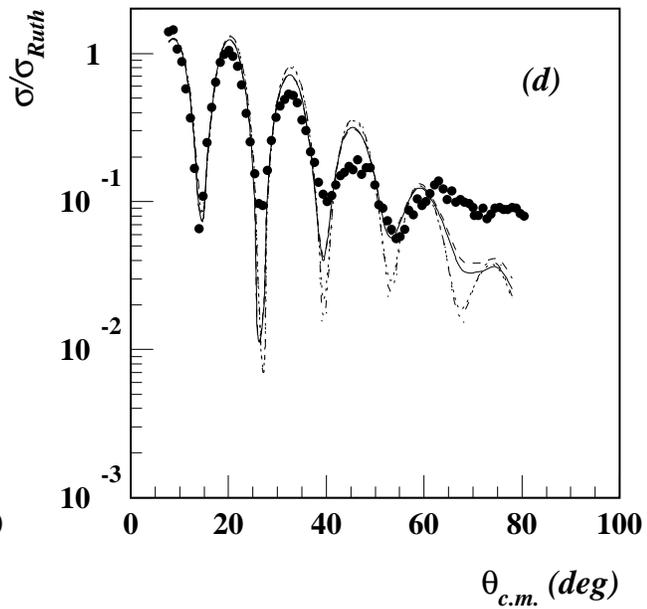
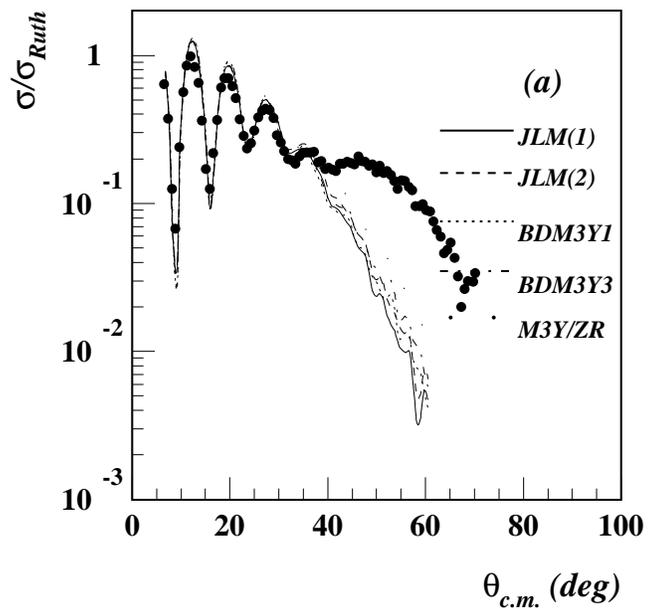
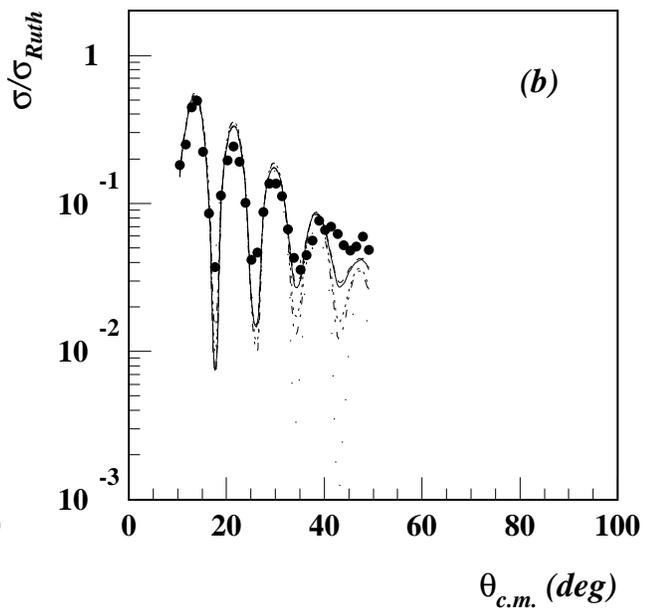
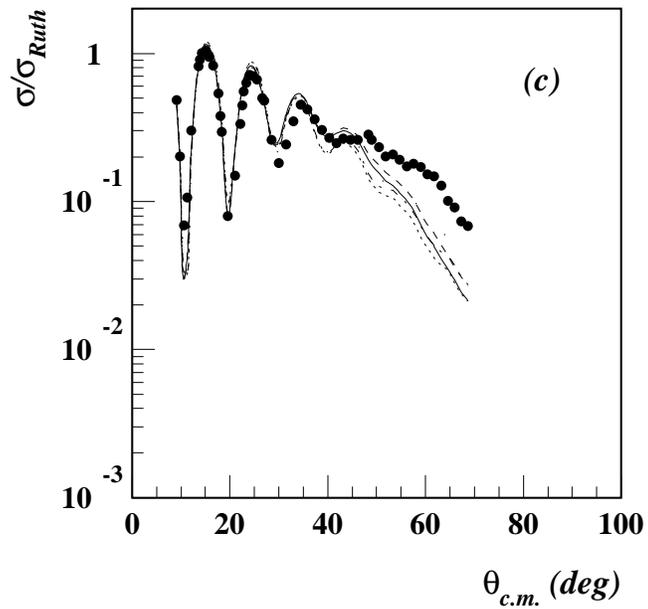
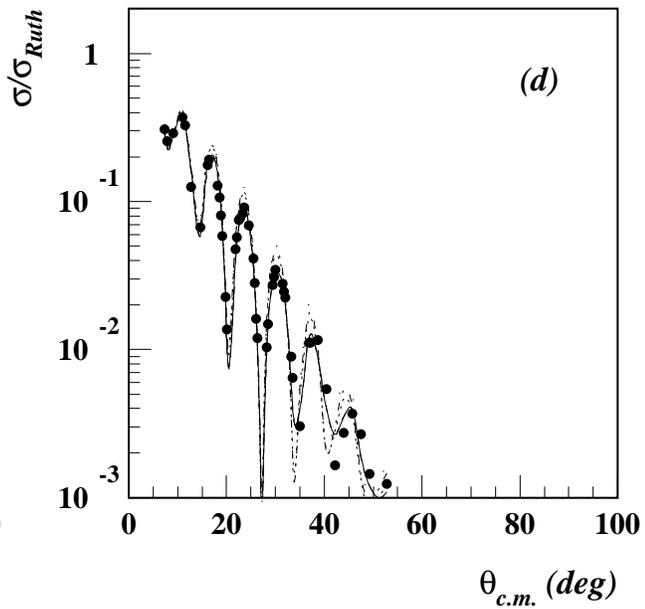
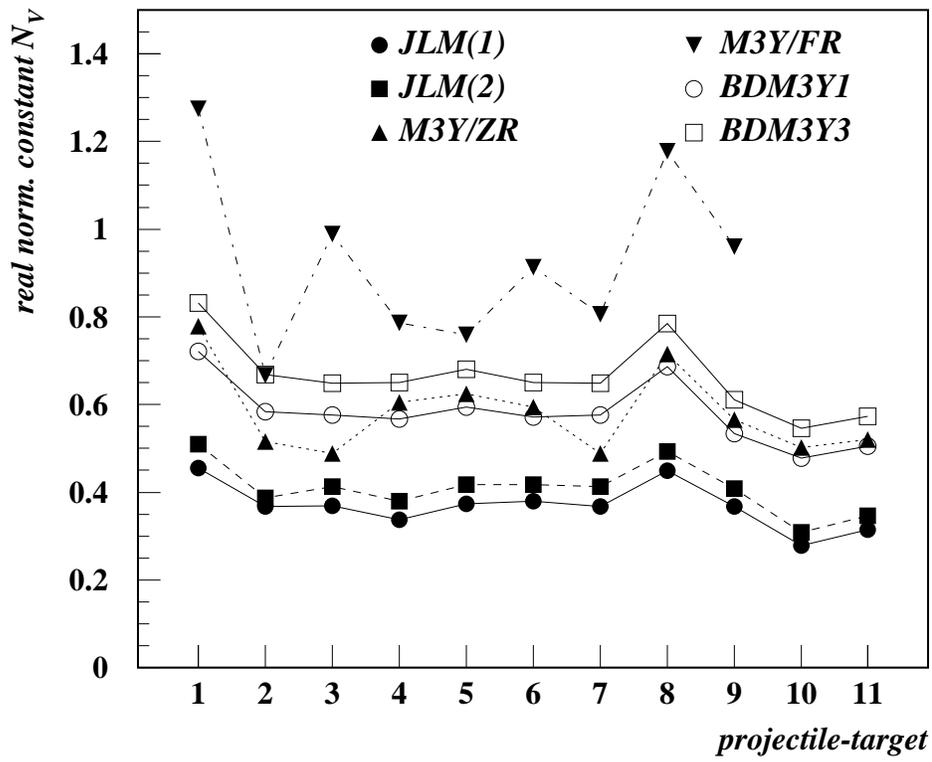
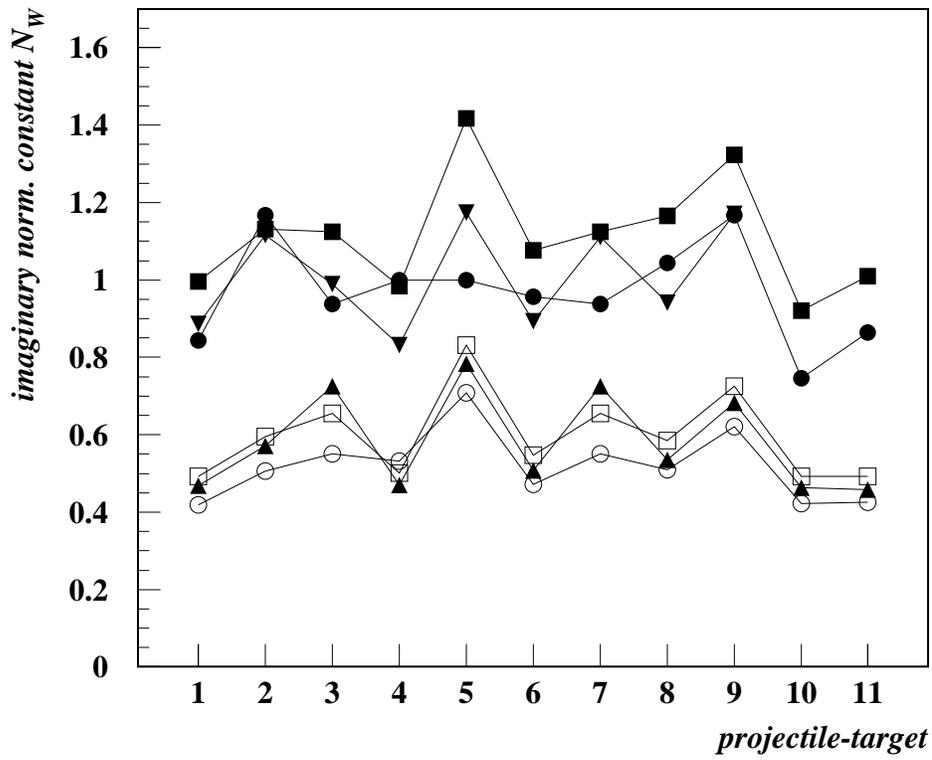
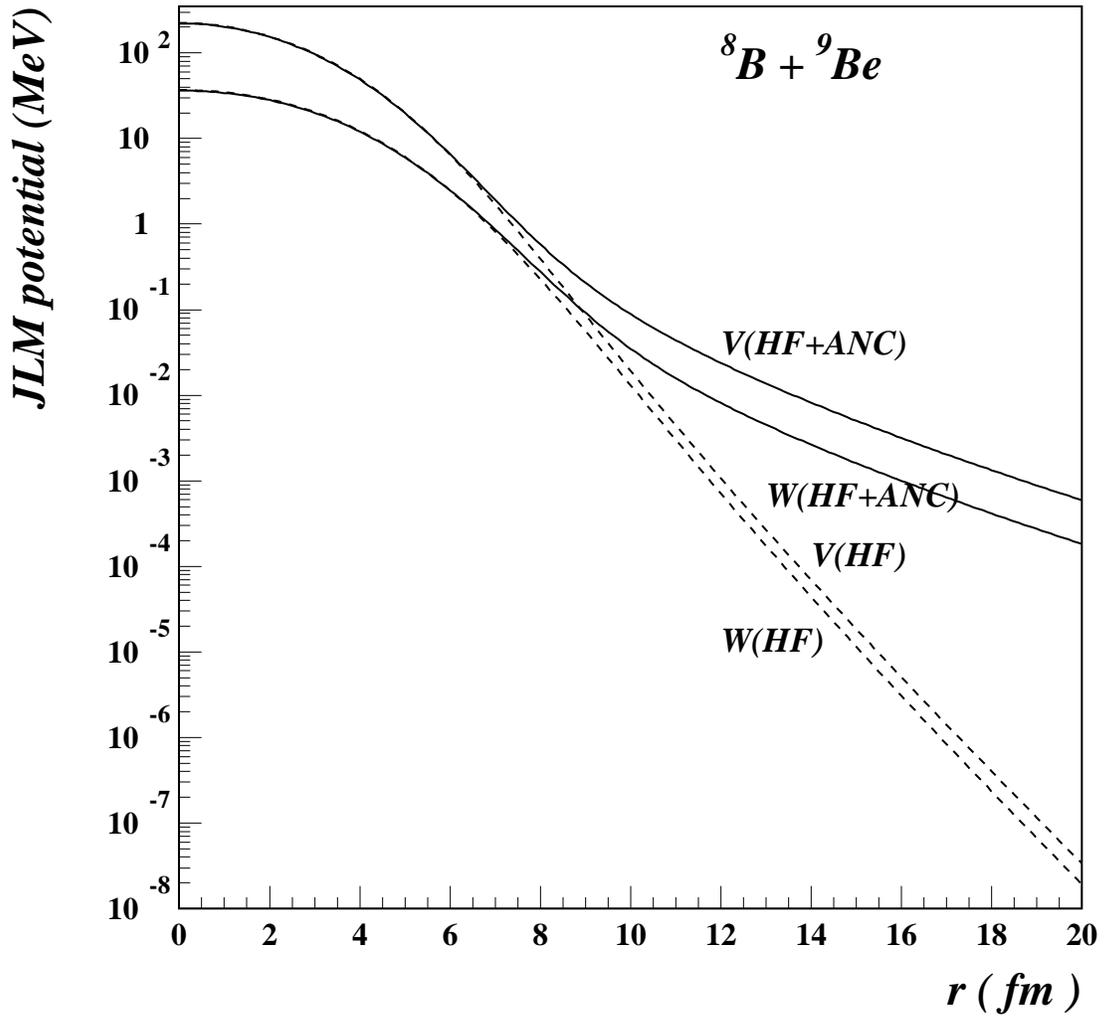
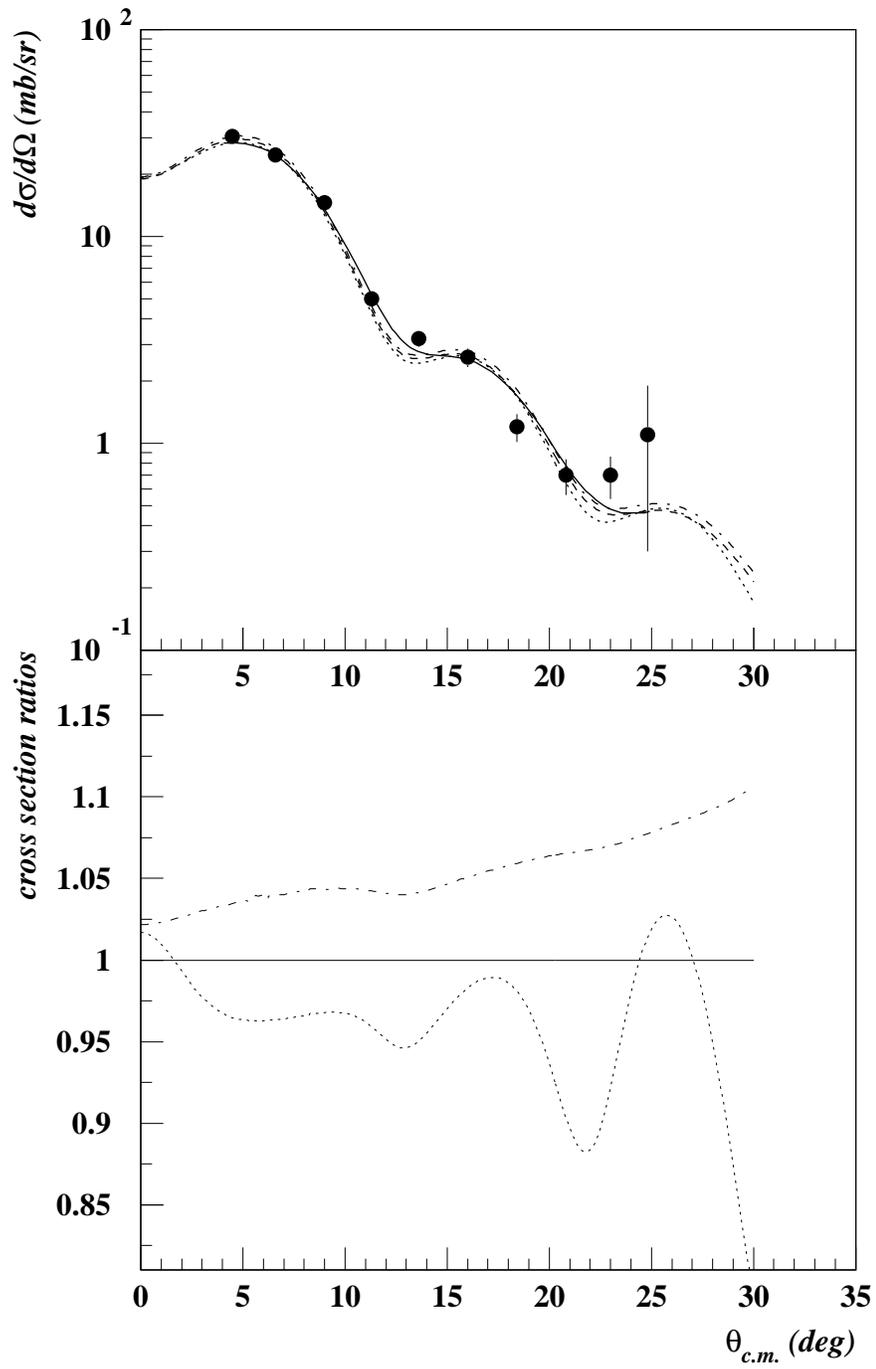
\end{document}